\documentclass[12pt]{article}

\usepackage{amsmath,graphicx}
\usepackage{epstopdf}
\usepackage{multirow}
\usepackage{amsfonts}
\usepackage{amssymb}
\usepackage{amscd}

\def\hybrid{\topmargin -20pt    \oddsidemargin 0pt
        \headheight 0pt \headsep 0pt
        \textwidth 6.25in       
        \textheight 9.5in       
        \marginparwidth .875in
        \parskip 5pt plus 1pt   \jot = 1.5ex}

\hybrid

\numberwithin{equation}{section}
\numberwithin{table}{section}

\newcommand{\beq}{\begin{equation}\begin{aligned}}
\newcommand{\eeq}{\end{aligned}\end{equation}}
\newcommand{\bi}{\begin{itemize}}
\newcommand{\ei}{\end{itemize}}
\newcommand{\bea}{\begin{eqnarray}}
\newcommand{\eea}{\end{eqnarray}}
\newcommand{\ba}{\begin{array}}
\newcommand{\ea}{\end{array}}
\newcommand{\bt}{\begin{tabular}}
\newcommand{\et}{\end{tabular}}
\newcommand{\bc}{\begin{center}}
\newcommand{\ec}{\end{center}}

\newcommand{\dstyle}{\displaystyle}

\newcommand{\Gx}{\Gamma}

\newcommand{\cC}{A}
\newcommand{\cD}{\mathcal{D}}

\newcommand{\cF}{\mathcal{F}}

\newcommand{\cQ}{\mathcal{Q}}

\newcommand{\fs}{D}
\newcommand{\fst}{\tilde D}

\newcommand{\half}{\frac12}

\newcommand{\nn}{\nonumber}
\newcommand{\IM}{\textrm{Im} \,}
\newcommand{\RE}{\textrm{Re} \,}

\newcommand{\cref}{{\bf [check ref]}}

\newcommand{\N}{\mathcal{N}}

\newcommand{\phione}{{\alpha}}
\newcommand{\phitwo}{{\beta}}
\newcommand{\m}{{p}}
\newcommand{\p}{{m}}
\newcommand{\df}{\dd_4}
\newcommand{\Ae}{\Sigma^+}
\newcommand{\Ao}{\Sigma^-}
\newcommand{\SSe}{{\mathcal{S}_+}}
\newcommand{\SSo}{{\mathcal{S}_-}}

\newcommand{\del}{\partial}

\def\stt {$SU(3) \times SU(3)$}

\newcommand{\pure}{\Phi}
\newcommand{\hA}{\hat A}
\newcommand{\hI}{\hat I}

\newcommand{\id}{\mathbf{1}}

\newcommand{\cP}{\mathcal{P}}

\newcommand{\dd}{\mathrm{d}}
\newcommand{\ee}{\mathrm{e}}
\newcommand{\ii}{\mathrm{i}}

\newcommand{\der}{\partial}

\newcommand{\bbZ}{\mathbb{Z}}
\newcommand{\bbR}{\mathbb{R}}
\newcommand{\bbC}{\mathbb{C}}

\DeclareMathOperator{\GL}{\mathit{GL}}
\DeclareMathOperator{\SU}{\mathit{SU}}
\DeclareMathOperator{\SO}{\mathit{SO}}
\DeclareMathOperator{\Symp}{\mathit{Sp}}
\DeclareMathOperator{\Spin}{\mathit{Spin}}

\DeclareMathOperator{\img}{im}

\DeclareMathOperator{\Cliff}{Cliff}

\newcommand{\rep}[1]{\mathbf{#1}}

\DeclareMathOperator{\tr}{tr}

\DeclareMathOperator{\re}{Re}
\DeclareMathOperator{\im}{Im}

\newcommand{\mukai}[2]{\big<{#1},{#2}\big>}
\newcommand{\revmukai}[2]{\left<{#2},{#1}\right>}

\newcommand{\Lodd}{\Lambda^\text{odd}}
\newcommand{\Leven}{\Lambda^\text{even}}
\newcommand{\Leo}{\Lambda^\text{even/odd}}

\newcommand{\chis}{\chi_\epsilon}

\newcommand{\psis}{\psi_\epsilon}

\newcommand{\Bfl}{B^\text{fl}}
\newcommand{\Bdel}{\tilde{B}}

\newcommand{\vt}{U}

\newcommand{\beve}{b^+}
\newcommand{\bodd}{b^-}

\newcommand{\Gv}{\check{\Gamma}}
\newcommand{\Gf}{\hat{\Gamma}}

\newcommand{\scs}[1]{C^\infty(#1)}
\newcommand{\tn}{\sigma}

\newcommand{\omo}{\omega^{(0)}}

\newcommand{\cycle}{\text{cyclic}}

\begin{document}


\begin{titlepage}
\begin{center}

\rightline{\small SPhT-T06/166}
\rightline{\small ZMP-HH/06-19}
\rightline{\small Imperial/TP/06/DW/01}
\vfill

{\Large \bf $\SU(3)\times \SU(3)$ compactification \\*[5pt] 
  and mirror duals of magnetic fluxes}

\vskip 0.6cm

{\bf Mariana Gra{\~n}a$^{a}$, Jan Louis$^{b}$ and
Daniel Waldram$^{c}$}

\vskip 0.6cm

\begin{small}
{}$^{a}${\em Service de Physique Th\'eorique,                   
CEA/ Saclay \\
91191 Gif-sur-Yvette Cedex, France}  \\
\vskip 0.18cm
{\tt mariana.grana@cea.fr}
\vskip 0.4cm

{}$^{b}${\em II. Institut f{\"u}r Theoretische Physik der
  Universit{\"a}t Hamburg\\ 
Luruper Chaussee 149,  D-22761 Hamburg, Germany.}\\
\vskip 0.18cm
{\em Zentrum f\"ur Mathematische Physik, 
Universit\"at Hamburg,\\
Bundesstrasse 55, D-20146 Hamburg, Germany.}\\
\vskip 0.18cm
{\tt jan.louis@desy.de} 
\vskip 0.4cm

{}$^{c}${\em Blackett Laboratory, Imperial College London\\
London, SW7 2BZ, U.K.}\\
\vskip 0.18cm
{\em Institute for Mathematical Sciences, 
  Imperial College London\\
London, SW7 2PG, U.K.}\\
\vskip 0.18cm
{\tt d.waldram@imperial.ac.uk}
\end{small}
\end{center}

\vskip 0.6cm

\begin{center} {\bf Abstract } \end{center}
\vspace{-4mm}

\noindent
This paper analyses type II string theories in backgrounds which admit an
$\SU(3)\times \SU(3)$ structure. Such backgrounds are designed to
linearly realize eight out of the original 32 supercharges and as a
consequence the low-energy effective action can be written in terms of
couplings which are closely related to the couplings of
four-dimensional $N=2$ theories. This generalizes the previously 
studied case of $\SU(3)$ backgrounds in that the left- and right-moving sector
each have a different globally defined spinor. Given a truncation to a
finite number of modes, these backgrounds lead to a conventional
four-dimensional low-energy effective theory. The results are manifestly
mirror symmetric and give terms corresponding to the mirror
dual couplings of Calabi-Yau compactifications with magnetic
fluxes. It is argued, however, that generically such backgrounds are
non-geometric and hence the supergravity analysis is 
not strictly valid. Remarkably, the naive generalization of the
geometrical expressions nonetheless appears to give the correct
low-energy effective theory.  

\vskip 0.4cm
\noindent December 2006

\end{titlepage}

\tableofcontents


\section{Introduction}


String backgrounds which include non-trivial fluxes and are described
by generalized geometry have been of considerable interest
recently~\cite{grana}. The primary reason is that such generalized
compactifications are necessary whenever the string background
contains D-branes. Generalized
geometries have also featured prominently in recent mathematical
investigations since they provide interesting extensions of certain
established geometrical concepts such as complex and symplectic
geometry~\cite{3form}--\cite{Witt}.  

A particular aspect of generalized geometries is that they can appear
as mirror partners of Calabi--Yau compactifications with background
fluxes~\cite{GLMW}--\cite{Tomasiello} or as non-perturbative duals of
heterotic flux compactifications \cite{LM2}. 
More specifically, if one considers
type IIB supergravity compactified on Calabi-Yau threefolds one can
turn on non-trivial three-form flux for both the Ramond-Ramond (RR)
three-form $F_3$ and the Neveu-Schwarz (NS) three-form
$H_3$. In the mirror symmetric type IIA background the RR three-form
flux is mapped to RR-flux of the even field strength
$F^+=F_0+F_2+F_4+F_6$~\cite{GVW,LM}. On the other hand the NS
three-form flux becomes part of the geometry in the mirror dual
compactification~\cite{Vafa,GLMW}. More precisely, a Calabi-Yau
compactification with electric NS three-form flux is conjectured to be
mirror symmetric to compactifications on manifolds known as
``half-flat manifolds''~\cite{HitchinHF,CS,GLMW}.\footnote{The notion 
  of electric flux is related to the definition of the Abelian
  (electric) gauge bosons. In type IIB they arise from expanding the
  RR four-form $C_4$ in terms of elements of the third cohomology
  $H^3$ of the Calabi-Yau. On $H^3$ there is a natural symplectic
  structure which in physical terms can be used to define electric
  gauge bosons and their magnetic duals. With this definition in mind
  one also has a natural split of the NS  three-form flux into
  electric and magnetic. (See~\cite{LM,GLMW} for further details.)} 
These six-dimensional manifolds are a specific subclass
of manifolds with $\SU(3)$ structure. A generic 
manifold  with $\SU(3)$ structure admits a nowhere vanishing, globally
defined spinor $\eta$ which, however, is not necessarily covariantly
constant with respect to the Levi-Civita connection. 
In
this sense manifolds with $\SU(3)$ structure generalize the notion of 
Calabi-Yau manifolds.\footnote{In the context of string theory such
manifolds were first discussed in \cite{Rocek}.}

The mirror of Calabi-Yau compactifications with 
magnetic three-form fluxes turns out 
to be more involved. The types of gaugings 
arising in such compactifications were discussed in 
\cite{DFTV}.
In refs.~\cite{MR,Hull,STW} it has been suggested 
that the corresponding mirror backgrounds do not correspond to 
conventional geometric compactifications. Such non-geometrical 
backgrounds have been studied from different points of view
in refs.~\cite{MR}--\cite{IE}. In ref.~\cite{GLW} we conjectured that
the mirror of 
the magnetic fluxes is found among compactifications on manifolds with  
$\SU(3)\times \SU(3)$ structure~\cite{Gualtieri,JW,GMPT2}. Such
manifolds are generalizations of manifolds with $\SU(3)$ structure in
that they admit two globally defined spinors, one for each of the two
original ten-dimensional supersymmetries. Recently the relationship
between these different proposals has been clarified in
ref.~\cite{pascalsakura}. For $N=1$ orientifold
compactification our proposal for mirror symmetry was indeed confirmed
in ref.~\cite{iman}. Mirror symmetry can also be discussed in terms of
brane configurations, which in this context are naturally described by
calibrations in generalised geometry~\cite{Gualtieri,gencal}. 

In ref.~\cite{GLW} we showed that compactifications on manifolds with
$\SU(3)\times \SU(3)$ structure are the most general geometric
compactifications of type II supergravities with eight unbroken
supercharges or,  from a four-dimensional point of view, with 
$N=2$ supersymmetry. The corresponding low-energy effective action
depends only on the light modes of the string while the heavy string-
and Kaluza-Klein excitations are integrated out. The couplings of this
action are strongly constrained by the unbroken $N=2$ supersymmetry
which leads to an intricate interplay between supersymmetry and
geometry. For generalized compactifications the distinction between
heavy and light modes is not straightforward and as a consequence the
definition of the effective action is somewhat
ambiguous. In~\cite{GLW} we showed that even without any Kaluza-Klein
truncation it is possible to rewrite the ten-dimensional effective
action in a background with $\SU(3)$-structure in a form which
linearly realizes the eight unbroken supersymmetries. Or in other
words we defined the equivalent of the standard $N=2$ couplings, that
is the holomorphic prepotential and the Killing prepotentials, but now 
in ten dimensions and showed that they do obey the constraints of
$N=2$ supersymmetry. From a four-dimensional point of view this action
contains an infinite number of modes and a Kaluza-Klein reduction then
corresponds to a consistent truncation to a finite subspace.  

The purpose of this paper is to fill in two missing elements of our earlier
work. We first reanalyze part of the reformulation of ten-dimensional
type II supergravity in terms of Hitchin's generalized geometrical
structures given in~\cite{GLW}. Specifically we derive the form
of the Killing prepotentials (the $N=2$ analogue of the
superpotential and $D$-terms) in the case of a generic
$\SU(3)\times\SU(3)$ structure, verifying the expressions conjectured
in~\cite{GLW}. We then discuss the truncation to a finite set of
modes, leading to a conventional four-dimensional effective theory. In
this paper we do not address directly the question of when such
truncations exist, but simply derive a set of consistency conditions
for the effective theory to be $N=2$ supersymmetric. (These issues are
discussed in detail in~\cite{mkp}.) Given such a truncation, we
identify the backgrounds mirror to a Calabi--Yau compactification with
magnetic $H$-flux, the case which was missing from the analysis
of~\cite{GLMW}. We then use existing work to argue that generically
these are in fact non-geometrical. Nonetheless, the corresponding
low-energy effective theories can be derived from the  
general \stt\ structure expressions, given some suitable truncation,
despite the fact that these were derived assuming there was a
geometrical compactification. This is consistent with the fact that at
least some of the non-geometrical backgrounds are geometrical on any
local patch. 

The structure of the paper is as follows. In section~\ref{structure}
we review the geometry of generalized structures and show how they can be
used to rewrite type II sypergravity in a form analogous to $d=4$,
$N=2$ supergravity. In section~\ref{projection}, we show in detail how
the spectrum of the supergravity fluctuations can be arranged into
$N=2$ -- like multiplets and in addition, what representations need to be
projected out in order to define a theory without additional
spin-$\tfrac{3}{2}$ multiplets. In section~\ref{superpotential} we
derive the analogs of the Killing prepotentials for the generic
theory, verifying the form conjectured in~\cite{GLW}. In
section~\ref{magnetic} we  show that one can identify a specific \stt\
structure with an appropriate mode expansion
of the supergravity fields which reproduces the mirror dual
low-energy effective theory  of  Calabi--Yau compactifications with magnetic
$H$-flux. In section~\ref{general} we consider 
generic  \stt\ structures and compute the Killing prepotentials of the
corresponding compactified type IIA and type IIB theories. 
They turn out to be  manifestly mirror symmetric and
all known compactifications can be obtained from them as special
cases.\footnote{A specific set of generalized mirror manifolds has
been constructed in \cite{Chuang:2005qd}.}
In section~7 we take up the issue of non-geometric compactifications
and show that backgrounds with \stt\ structure generically also
contain non-geometric backgrounds.
Finally, section~\ref{conclusion} concludes with some open problems.
Our conventions for $\Spin(6)$ and $\Spin(6,6)$ spinors are given in
Appendix~\ref{app:spinors} while the conditions for a consistent mode
truncation are spelled out in Appendix~\ref{app:expand}.


\section{Supergravity and $\SU(3)\times\SU(3)$ structures}
\label{structure}


We begin by briefly reviewing the reformulation of ten-dimensional
type II supergravities given in~\cite{GLW}
and some of the key ingredients of generalized geometry in six
dimensions. Recall that supersymmetry variations in type II
supergravity are given by a pair of ten-dimensional spinors
$(\epsilon^1,\epsilon^2)$. In the reformulation, we concentrate on an
eight-dimensional subset of supersymmetries, analogous to the eight
supersymmetries of $N=2$ supergravity in four ($d=4$) space-time
dimensions. Since there are no eight-dimensional representations of
$\Spin(9,1)$, this rewriting necessarily no longer has manifest
ten-dimensional Lorentz symmetry, but, as we will see, the bosonic
fields can actually be arranged in terms of $O(6,6)$ representations
which are the natural objects describing generalized geometry.  

Specifically, decomposing $\Spin(9,1)$ into $\Spin(3,1)\times\Spin(6)$
subgroups we identify eight supersymmetry parameters given by 
\begin{equation}
\begin{aligned}
\label{decompepsilon}
   \epsilon^1 &= \varepsilon_+^1 \otimes \eta^1_-
      + \varepsilon_-^1 \otimes \eta^1_+ \ , \\
   \epsilon^2 &= \varepsilon_+^2 \otimes \eta^2_\pm 
      + \varepsilon_-^2 \otimes \eta^2_\mp \ ,
\end{aligned}
\end{equation}
where in the second line we take the upper sign for type IIA and the
lower for type IIB. Here $\eta^A_+$ with $A=1,2$ are 
spinors of $\Spin(6)$ while $\varepsilon^A$ are Weyl spinors of
$\Spin(3,1)$. In each case
$\eta^A_-$ and $\varepsilon^A_-$ are the charge conjugate spinors and
the $\pm$ subscripts denote the chirality (for more details see
appendix~\ref{app:spinors}). For a given pair $(\eta_+^1,\eta_+^2)$ we
have eight spinors parametrized by $\varepsilon^A_\pm$. 
These are the eight supersymmetries which remain
manifest in the reformulated theory. 

The assumption that we can identify $\eta^A_+$ globally puts a
topological constraint on the ten-dimensional spacetime: it must admit
a pair of $\SU(3)$ structures, one for each spinor. The
tangent bundle must split according to $TM^{9,1}=T^{3,1}\oplus F$, where
$F$ admits a pair of nowhere vanishing spinors. 
A simple example of such a split is a space-time which is a product
$M^{9,1}=M^{3,1}\times M^{6}$ (with $M^6$ admiting two such spinors)
but the background under consideration
can also be more general.
The split of the tanget space implies
that all fields of the theory can be decomposed under
$\Spin(3,1)\times\Spin(6)$.  

The two spinors $\eta^A_+$ are not necessarily different. If they
coincide on the whole manifold, the two $\SU(3)$ structures are the
same, and the manifold has a single $\SU(3)$ structure. In
neighborhoods where the spinors are not parallel, two real vectors $v$
and $v'$  can be defined by the bilinear $v^m -i v'^m= \bar
\eta^1_+\gamma^m \eta^2_-$. If the spinors never coincide, this
complex vector is nowhere vanishing, and the two $\SU(3)$ structures
intersect globally in an $\SU(2)$ structure.  

Instead of defining a general $\SU(3)$ structure via the spinor $\eta$
one can equivalently define it by a real 
fundamental two-form $J$ and a complex three-form $\Omega$.
Analogously, a pair of  $\SU(3)$ structures  can be defined by a pair
 $(J^A, \Omega^A)$ which locally (in neighborhoods where the
 two structures do not coincide) can be given as \cite{waldram}
\begin{equation}\label{localSU2}
\begin{aligned}
   J^1 &= j + v \wedge v'\ , \qquad&
   \Omega^1 &= \omega \wedge(v+\ii v')\ , \\
   J^2 &= j - v \wedge v'\ , \qquad &
   \Omega^2 &= \omega \wedge(v-\ii v')\ .
\end{aligned}
\end{equation}
$v, v'$ are one-forms,
$j$ is a real two-form and $\omega$ is a complex two-form. 
Together $(j,\omega,v,v')$ define a local
$\SU(2)$ structure on $F$, if none of them has zeros they define a
global $\SU(2)$ structure.

Crucially, one finds, following Hitchin \cite{GCY,Gualtieri,JW}, that
the pair of $\SU(3)$ structures is actually better viewed as an
$\SU(3)\times\SU(3)$ structure on the generalized tangent bundle, that
is $F\oplus F^*$. In turn, this structure is defined by a pair of
$O(6,6)$ spinors. As a consequence, the bosonic supergravity fields
can then all be written in terms of $O(6,6)$ representations. To
briefly see how this works, let us start by recalling  some facts
about generalized geometry in six dimensions.  

There is a natural $O(6,6)$ metric on $F\oplus F^*$ given by
\begin{equation}
\label{metric}
   (V,V') = \tfrac{1}{2}i_x \xi'+ \tfrac{1}{2}i_{x'} \xi .
\end{equation}
where
$V=x+\xi, V'=x'+\xi' \in F\oplus F^*$. In a coordinate basis the
metric reads 
\begin{equation}
\label{sGdef}
   {\cal G} = \frac{1}{2}\left( 
      \begin{array}{cc} 0 & \Bbb I_6 \\ \Bbb I_6 & 0
      \end{array} \right)\ .
\end{equation}
Given this metric one can define $O(6,6)$ spinors. These are discussed
in detail in the appendix~\ref{app:spinors}, here we will summarize
some key points. It turns out that the spinor bundle $S$ is isomorphic
to the bundle of forms  
\begin{equation}
\label{spinbundles}
   S \simeq \Lambda^* F^*\ .
\end{equation}
Spinors of $O(6,6)$ can be chosen to be Majorana --Weyl. The positive
and negative helicity spin bundles $S^\pm$ are isomorphic to the  
bundles of even and odd forms $\Leo F^*$. The Clifford action on
$\chi\in\Lambda^*F^*$ is given by 
\begin{equation}
\label{cliff}
   (x+\xi)\cdot\chi = i_x\chi + \xi\wedge\chi\ .
\end{equation}
The isomorphism~\eqref{spinbundles} is not unique but is given by a
choice of volume form $\epsilon$ (though is independent of the sign of
$\epsilon$) \footnote{We are using the same symbol $\epsilon$ to
  denote the volume form and the ten-dimensional spinors. The
  distinction between the two should hopefully be clear given the
  context.} If $\chi\in\Lambda^*F^*$ we write $\chis\in S$ for the
corresponding spinor. The usual spinor bilinear form 
$\psis^t\cdot\chis$ on $S$ is then related to the Mukai pairing 
$\mukai{\cdot}{\cdot}$ on forms by 
\begin{equation}
\label{mukai}
   (\psis^t\cdot\chis)\,\epsilon
     = \mukai{\psi}{\chi} 
     = \sum_p (-)^{[(p+1)/2]} \psi_p \wedge \chi_{6-p}\ ,
\end{equation}
where the subscripts denote the degree of the component forms in
$\Lambda^*F^*$ and $[(p+1)/2]$ takes the
integer part of $(p+1)/2$.

A metric $g$ and $B$-field on $F$ naturally define an $O(6)\times
O(6)$ subgroup of $O(6,6)$ and hence a decomposition of $S$ into
$\Spin(6)$-bundles $S=S_1\otimes S_2$. The two $\Spin(6)$-spinors
$\eta^1_\pm$ and $\eta^2_\pm$ defined in~\eqref{decompepsilon} are
naturally sections of $S_1$ and $S_2$ respectively. In terms of the
diagonal $\Spin(6)$ group under which we identify $S_1\simeq S_2$, we
can view $\chis\in S$ as a $\Spin(6)$ bispinor, that is, as an element
of $\Cliff(6,0;\bbR)$. Explicitly one can write real $\chis^\pm\in
S^\pm$ as 
\begin{equation}
\label{66decomp}
   \chi_\epsilon^\pm = \zeta_+\bar{\zeta}'_\pm 
      \pm  \zeta_-\bar{\zeta}'_\mp\ ,
\end{equation}
where $\zeta_+$, $\zeta'_+$ are ordinary $\Spin(6)$ spinors and
elements of $S_1^+$ and $S_2^+$ respectively. From this perspective
$\chis^\pm$ is a matrix. In fact it can be expanded as  
\begin{equation}
\label{chichiepsilon}
   \chis^\pm = \sum_p
      \frac{1}{8p!}\chi^\pm_{m_1\dots m_p} \gamma^{m_1\dots m_p} , 
\end{equation}
with
\begin{equation}
   \chi^\pm_{m_1\dots m_p} = \tr (\chi^\pm \gamma_{m_p\dots m_1})
      \in \Lambda^pF ,
\end{equation}
and where $\gamma^m$ are $\Spin(6)$ gamma-matrices and the trace is
over the $\Spin(6)$ indices. For $\chi_\epsilon^+$ only the even forms
are non-zero, while for $\chi_\epsilon^-$ the odd forms are
non-zero. This gives an explicit realisation of the isomorphism
between $S^\pm$ and $\Leo F^*$ using the volume form $\epsilon_g$
compatible with the metric $g$. 

Explicitly the $O(6,6)$ Clifford action~\eqref{cliff} is realised in
terms of commutators and anticommutators 
\begin{equation}
\label{Gdecomp}
   (x+\xi)\cdot\chi_\epsilon^\pm 
      = \tfrac{1}{2}[ x^m\gamma_m , \chi_\epsilon^\pm ]_\mp 
         + \tfrac{1}{2}[ \xi_m\gamma^m , \chi_\epsilon^\pm ]_\pm\ . 
\end{equation}
Similarly the Mukai pairing is given by 
\begin{equation}
\label{mukaispinors}
   \mukai{\psi}{\chi} 
      = - 8\tr(\psis^t\chis)\epsilon_g\ . 
\end{equation}
where 
\begin{equation}
   \psis^t := \gamma_{(6)} C \psis^T C^{-1}\ ,
\end{equation}
with $\gamma^{(6)}=\frac{1}{6!}\epsilon_g^{m_1\dots
  m_6}\gamma_{m_1\dots m_6}$ and $\epsilon_g$ is the natural
orientation compatible with the metric $g$ (defined up to an arbitrary
sign). (See Appendix \ref{app:spinors} for more details.)
 
Now consider the pair of complex $O(6,6)$ spinors 
\begin{equation}
\begin{aligned}
\label{purespinors}
   \Phi^+ &= \ee^{-B} \Phi^+_0 := \ee^{-B} \eta^1_+\bar{\eta}^2_+ \ , \\
   \Phi^- &= \ee^{-B} \Phi^-_0 := \ee^{-B} \eta^1_+\bar{\eta}^2_- \ ,
\end{aligned}
\end{equation}
where $B$ is the NS two-form on $F$ and $\ee^{-B}$ acts by wedge
product. First one notes that when $B$ is non-trivial, $\Phi^\pm$ are
actually not quite sections of $S^\pm$. Instead one must consider the
extension $E$  
\begin{equation}
\label{extension}
   0 \longrightarrow F^* \longrightarrow E \longrightarrow F
   \longrightarrow 0\ ,
\end{equation}
defined as follows. If on the overlap of two patches $U_\alpha\cap
U_\beta$ the $B$-field is patched by 
\begin{equation}
   B_\alpha = B_\beta + \dd A_{\alpha\beta} 
\end{equation}
then in the extension~\eqref{extension} we must identify 
\begin{equation} \label{patching}
   x_\alpha + \xi_\alpha = x_\beta + 
      \left(\xi_\beta + i_{x_\beta}\dd A_{\alpha\beta}\right) .
\end{equation}
Since $i_{x_\alpha}\xi_\alpha=i_{x_\beta}\xi_\beta$, the $O(d,d)$
metric can still be defined on the extension $E$ and thus one can define
spinor bundles $S^\pm(E)$ and hence $\Phi^\pm\in S^\pm(E)$.

In order to introduce the notion of pure spinors we need to define 
the anihilator space $L_\Phi$ of an $O(6,6)$ spinor as 
\beq
L_\Phi=\{ V\in E: V \cdot \Phi =0\}\ .
\eeq
A spinor is called pure whenever its annihilator space is maximal
isotropic, that is  $L_\Phi$ is six-dimensional, and $\forall \, V,V'
\in L_\Phi$, $(V,V')=0$ holds. A pure spinor $\Phi$ therefore induces
a decomposition $E=L_\Phi+\bar L_\Phi$.  
The complex $O(6,6)$ spinors $\Phi^\pm$ defined in (\ref{purespinors})
are pure spinors.

Individually $\Phi^\pm$ each defines an $\SU(3,3)$ structure on
$E$. Provided these structures are compatible, together they define a
common $\SU(3)\times\SU(3)$ structure. The requirements of
compatibility are that $\mathrm{dim}(L_{\Phi^+} \cap L_{\Phi^-})=3$, and
that $\Phi^\pm$ have the same normalization \cite{Gualtieri}. In terms
of Mukai pairings, they read \cite{GLW}
\begin{eqnarray}
   \mukai{\Phi^+}{V\cdot\Phi^-}& =&  \mukai{\bar
     \Phi^+}{V\cdot\Phi^-}=0  \quad \forall \, V\in E\
   , \label{comp1}\\ 
   \mukai{\Phi^+}{\bar{\Phi}^+} &=& \mukai{\Phi^-}{\bar{\Phi}^-}\
   . \label{comp2} 
\end{eqnarray}

If $\Phi^\pm$ are built out of $Spin(6)$ spinors in the form of Eq. (\ref{purespinors}), they
are automatically compatible \cite{GMPT2}.
The pair $\Phi^\pm$ in (\ref{purespinors}) therefore defines 
an $\SU(3)\times\SU(3)$ structure on
$E$. In particular, one can see that they are invariant under
independent $\SU(3)$ groups acting on $\eta^1$ and $\eta^2$. Note that
in terms of the local $\SU(2)$ structure \eqref{localSU2} they are 
given by \cite{JW,GMPT2}
\begin{equation}
\label{exppurespinors}
\begin{aligned}
   \Phi^+ &= \ee^{-B} \left(
      \bar c_{\|} \, \ee^{-\ii j} - \ii \bar c_{\bot} \omega \right) \wedge
      \ee^{-\ii v\wedge v'}\ , \\
   \Phi^- &= - \ee^{-B} \left( c_{\bot}
      \ee^{-\ii j} +  \ii c_{\|} \,\omega \right) \wedge
      (v+\ii v')\ ,
\end{aligned}
\end{equation}
where $c_\|, c_\bot$ are complex functions satisfying $|c_\||^2+|c_\bot|^2=1$. $c_\|$ ($c_\bot$)  vanishes when the two spinors $\eta^{1,2}$ are
orthogonal (parallel), namely $\eta^2_+=c_\| \, \eta^1_+ + c_\bot (v+iv')_m \gamma^m \eta^1_-$. 
At the points where the spinors are parallell ($c_\bot =0$),
the expression (\ref{exppurespinors}) should be understood 
as $\Phi^+= \ee^{-B} \, \ee^{-\ii J}$, $\Phi^-=  -\ii \ee^{-B} \,
\Omega$, where $J$ and $\Omega$ are the two- and 
three-form of the single SU(3) structure defined by the coinciding spinors.
In this case, $\Phi^+$ defines a symplectic structure, and $\Phi^-$ a complex
structure. Complex and symplectic structures  are particular cases of generalized complex structures.
In this situation the compatibility conditions (\ref{comp1}) imply the
familiar requirements $J\wedge \Omega=0, B\wedge \Omega =0$ while the
normalization condition (\ref{comp2}) implies $J\wedge J\wedge
J=\tfrac34 \ii  \Omega \wedge \bar \Omega$.   
In the general case, $\Phi^-$ contains not only a 3-form, but also a 1 and a 
5-form, and defines a generalized complex structure that is not purely
complex but is a mixture of complex and symplectic structures. 

One key point in connecting these generalised geometrical structures
to supergravity, is that, following Hitchin~\cite{3form}-\cite{GCY},
one can show that there is a natural special K\"ahler structure on the
space of pure spinors at a point. Furthermore, this structure
precisely gives the metric for the ``four-dimensional'' kinetic terms
in the reformulation of type II supergravity in a $N=2$
four-dimensional-type form~\cite{GLW}. This structure is reviewed in the
appendix~\ref{app:expand}. The second key point is that the 
prepotentials, which describe the potential terms and gaugings of the
$N=2$ theory, are also naturally defined in terms of generalised
geometrical structures. This is discussed in
section~\ref{superpotential}. 

Here, let us first briefly summarize the special K\"ahler
structure. Working at a fixed point in the manifold, one starts with a
real stable $\Spin(6,6)$ spinor, or its associated form
$\chi^\pm$. Such form is stable if it lies in an open orbit of
$\Spin(6,6)$. 
One can construct a $\Spin(6,6)$ invariant six-form, known as the 
Hitchin function $H(\chi^\pm)$, which is homogeneous of degree two as a
function of $\chi^\pm$. One can get a second real form by derivation of
the Hitchin function: $\hat{\chi}^\pm(\chi):=-\del  H(\chi^\pm)/\del
\chi^\pm$. This form $\hat{\chi}^\pm$ has the same parity as
$\chi^{\pm}$, and can be used to define the complex spinors
$\Phi^{\pm}=\frac{1}{2}(\chi^\pm+\ii\hat{\chi}^\pm)$. Hitchin showed 
that the complex spinors built in this form are pure. Since $H$ is
homogeneous of degree two in $\chi^\pm$, we have 
\beq
H(\Phi^\pm) = \tfrac{1}{2} \mukai{\chi^\pm}{\hat\chi^\pm} = \ii
\mukai{\Phi^\pm}{\bar \Phi^\pm}  \ .
\eeq
There is a symplectic structure on the space of stable spinors given
by the Mukai pairing and a complex structure corresponding to the
complex spinor $\Phi^\pm$. Both complex and symplectic structures are
integrable, and therefore the space of stable forms (or pure spinors)
is K\"ahler, or rather it is rigid special K\"ahler (for more details,
see appendix~\ref{app:expand}). Quotenting this space by the $\mathbb C^*$
action $\Phi^\pm\to \lambda\Phi^\pm$ for $\lambda \in \mathbb C^*$
(i.e., modding out by rescalings of the pure spinor), gives a space
with a the K\"ahler potential $K$  is related to the Hitchin function
by 
\beq \label{Kahlerpot}
\ee^{-K^\pm} =  H(\Phi^\pm)= \ii  \mukai{\Phi^\pm}{\bar \Phi^\pm} \ ,
\eeq
which defines a local special K\"ahler metric. 

For a single $SU(3)$ structure, i.e. for
$\Phi^+=\ee^{-(B+\ii J)}$, 
$\Phi^-=-\ii\ee^{-B} \Omega$, the K\"ahler potentials
(\ref{Kahlerpot}) are given respectively by the familiar expressions
\begin{equation}
   \ee^{-K^+}=\tfrac{4}{3} J \wedge J \wedge J , \qquad \qquad 
   \ee^{-K^-}= \ii \Omega \wedge \bar \Omega. 
\end{equation}
Note that $B$ drops from these expressions (which is easy to see since 
$\mukai{\ee^{-B}\psi}{\ee^{-B}\chi}= 
\mukai{\psi}{\ee^{B}\ee^{-B}\chi}=\mukai{\chi}{\psi}$).  

In the following it will be useful to have a decomposition of $O(6,6)$
spinors under the \stt\ subgroup defined by $\Phi^+$ and
$\Phi^-$. From (\ref{66decomp}) the decomposition of a positive
chirality spinor under $\Spin(6)\times \Spin(6)$ is given by 
\beq
\rep{32}^+ = (\rep{4},\rep{4}) + (\bar{\rep{4}}, \bar{\rep{4}}) \ .
\eeq
Under each $\SU(3)$ subgroup of $\Spin(6)$ we have
$\rep{4}=\rep{1}+\rep{3}$. Hence under $\SU(3)\times\SU(3)$, the
$O(6,6)$ spinor decomposes into 8 different representations. A similar
decomposition of a negative chirality $O(6,6)$ spinor gives eight
further representations. Denoting by  $U_{\rep{r}, \rep{s}}$ the set
of forms transforming in the $(\rep{r}, \rep{s})$ representation of
\stt\, together these decompositions can be arranged in a diamond as
given in Table \ref{diamond} \cite{GHodge}.\footnote{By an abuse of
notation, it is 
  convenient to use $\rep{\bar{1}}$ to denote the singlet coming from
  the decomposition of $\rep{\bar{4}}$.} 
\begin{table}[htdp]
\begin{center}
\begin{tabular}{ccccccc}
&&& $U_{\rep{1},\rep{\bar 1}}$ &&& \\
&& $U_{\rep{\bar 3},\rep{\bar 1}}$ && $U_{\rep{1},\rep{3}} $ && \\
& $U_{\rep{3},\rep{\bar 1}}$ && $U_{\rep{\bar 3},\rep{3}}$ && $U_{\rep{1},\rep{\bar 3}}$ &\\
$U_{\rep{\bar 1},\rep{\bar 1}}$ && $U_{\rep{3},\rep{3}}$ && $U_{\rep{\bar 3},\rep{\bar 3}}$ && 
$U_{\rep{1},\rep{1}}$ \\
& $U_{\rep{\bar 1},\rep{ 3}}$ && $U_{\rep{3},\rep{\bar 3}}$ && $U_{\rep{\bar 3},\rep{1}}$ &\\
&& $U_{\rep{\bar 1},\rep{\bar 3}}$ && $U_{\rep{3},\rep{1}} $ && \\
&&& $U_{\rep{\bar 1},\rep{1}}$ &&& \\
 \end{tabular}
 \caption{\small 
\textit{Generalized \stt\ diamond.}}\label{diamond}
\end{center}
\end{table}  
$U_{\rep{1}, \rep{\bar 1}}$ contains a sum of even forms while
$U_{\rep{\bar 3},\rep{\bar 1}}$ and $U_{\rep{1},\rep{3}} $ contain a
sum of odd forms. Similarly, third row consists of even forms, the
forth of odd forms and so on. Note that, unlike the decomposition of
forms induced by a complex structure into $(p,q)$-forms,  
the elements of $U_{\rep{r}, \rep{s}}$ are not necessarily of fixed
degree. Instead  $U_{\rep{r}, \rep{s}}$ contains forms of mixed degree
which however are always even or odd. For example, for a single
$\SU(3)$ structure on $F$ (which is a particular case of an  \stt\
structure on $E$), a form belongs to the space $U_{\rep{1}, \rep{\bar 1}}$ 
if it is a multiple of $\ee^{-(B+\ii J)}$. Thus it indeed contains all
even 0-, 2-, 4- and 6-form. Conversely forms of fixed degree are
linear combinations of elements in different $U$'s. For example, a
zero-form is a linear combination of elements in $U_{\rep{1},
  \rep{\bar 1}}\oplus U_{\rep{\bar 1}, \rep{1}}\oplus U_{\rep{3},
  \rep{\bar 3}} \oplus U_{\rep{\bar 3}, \rep{3}}$. 


\section{Field decompositions and spectrum}
\label{projection}


In this section we discuss the group-theoretical properties
of the massless type II supergravities fields in a background
with a generalized tanget bundle $T^{3,1}\oplus F\oplus F^*$.
In particular we show how the fields assemble in $N=2$ -- like
multiplets.

If $F\oplus F^*$ admits an \stt\ structure all ten-dimensional fields can be
decomposed under $\Spin(3,1) \times \SU(3)\times SU(3)$.
In fact it is slightly simpler to first go to light-cone gauge and
discuss the decompostion under $\SO(2) \times \SU(3)\times SU(3)$
instead. In order to 
do so let us first recall the decomposition of the two 8-dimensional
inequivalent Majorana-Weyl  
representations $\rep{8_S}$ and $\rep{8_C}$ and the vector
representation $\rep{8_V}$ of $\SO(8)$ under 
$\SO(8)\to \SO(2)\times\SO(6)\to \SO(2)\times\SU(3)$.
One has \cite{JP}
\beq\label{decom8s}
\begin{aligned}
 \rep{8_S}\ &\to\ \rep{4_{\half}}\oplus \rep{\bar 4_{-\half}}\
\to\ \rep{1_{\half}}\oplus \rep{1_{-\half}}\oplus \rep{3_{\half}}\oplus \rep{\bar 3_{-\half}}
\ , \\
 \rep{8_C}\ &\to\ \rep{4_{-\half}}\oplus \rep{\bar 4_{\half}}\
\to\ \rep{1_{\half}}\oplus \rep{1_{-\half}}\oplus \rep{3_{-\half}}\oplus \rep{\bar 3_{\half}}
\ ,\\
\rep{8_V}\ &\to\ \rep{1_{1}}\oplus \rep{1_{-1}}\oplus \rep{6_{0}}\
\to\ \rep{1_{1}}\oplus \rep{1_{-1}}\oplus \rep{3_{0}}\oplus \rep{\bar 3_{0}}
\ .
\end{aligned}
\eeq
where the subscript denotes the helicity of $\SO(2)$.

Let us start with the decomposition of the fermions which arise 
in the (NS,R) and (R,NS) sector. More precisly, 
in type IIA the two gravitinos together with the two dilatinos are in the
$(\rep{8_S},\rep{8_V})$
and $(\rep{8_V},\rep{8_C})$ of $\SO(8)_L\times \SO(8)_R$ while 
in type
IIB they come in the  $(\rep{8_S},\rep{8_V})$ and $(\rep{8_V},\rep{8_S})$
representations. The decomposition of these representations under 
$\SO(8)_{L}\times \SO(8)_R\to \SO(2)\times\SU(3)_{L}\times\SU(3)_{R}$
yields\footnote{The $\SO(2)$ factor in the decomposition of $\SO(8)_L$
  and $\SO(8)_R$ is of course the same.}
\beq\label{decom88}
\begin{aligned}
(\rep{8_S},\rep{8_V})\ \to\ &
(\rep{1},\rep{1})_{ \pm {\rep{\frac32}},\pm\rep\half} 
 \oplus(\rep{3},\rep{1})_{\rep{\frac32},-\rep\half}\oplus(\rep{\bar 3},\rep{1})_{-\rep{\frac32},\rep\half}
\oplus(\rep{1},\rep{3})_{\pm \rep\half}\oplus(\rep{1},\rep{\bar
  3})_{\pm \rep\half}\\
&\ \oplus(\rep{3},\rep{3})_{\rep\half}\oplus(\rep{\bar 3},\rep{3})_{-\rep\half}
\oplus(\rep{3},\rep{\bar 3})_{\rep\half}\oplus(\rep{\bar 3},\rep{\bar 3})_{-\rep\half}
\ , \\
(\rep{8_V},\rep{8_S})\ \to\ &
(\rep{1},\rep{1})_{\pm \rep{\frac32},\pm\rep\half} 
 \oplus(\rep{3},\rep{1})_{\pm \rep\half}\oplus(\rep{\bar 3},\rep{1})_{\pm \rep\half}
\oplus(\rep{1},\rep{3})_{\rep{\frac32},-\rep\half}\oplus(\rep{1},\rep{\bar 3})_{-\rep{\frac32},\rep\half}\\
&\ \oplus(\rep{3},\rep{3})_{\rep\half}\oplus(\rep{\bar 3},\rep{3})_{\rep\half}
\oplus(\rep{3},\rep{\bar 3})_{-\rep\half}\oplus(\rep{\bar 3},\rep{\bar 3})_{-\rep\half}
\ , \\
(\rep{8_V},\rep{8_C})\ \to\ &
(\rep{1},\rep{1})_{\pm \rep{\frac32},\pm\rep\half} 
 \oplus(\rep{3},\rep{1})_{\pm \rep\half}\oplus(\rep{\bar
   3},\rep{1})_{\pm \rep\half}
\oplus(\rep{1},\rep{3})_{-\rep{\frac32},\rep\half}\oplus(\rep{1},\rep{\bar 3})_{\rep{\frac32},-\rep\half}\\
&\ \oplus(\rep{3},\rep{3})_{-\rep\half}\oplus(\rep{\bar 3},\rep{3})_{-\rep\half}
\oplus(\rep{3},\rep{\bar 3})_{\rep\half}\oplus(\rep{\bar 3},\rep{\bar 3})_{\rep\half}
\ .
\end{aligned}
\eeq

Exactly as we did in ref.\ \cite{GLW} we do not consider the most
general $N=2$ -- like supergravity but only keep two gravitinos in the
gravitational multiplet and project out all other (possibly massive) 
spin-$\tfrac{3}{2}$ multiplets. This ensures 
a `standard' N=2 theory with only the gravitational
multiplet plus possibly vector, tensor and hypermultiplets.
In this case the couplings in the low energy effective action 
are well known and highly
constrained by the $N=2$ supersymmetry. 

{}From \eqref{decom88} we learn 
that keeping only the two gravitinos 
of the gravitational multiplet is insured if 
all  representations of the form 
$(\rep{3},\rep{1})$, $(\rep{\bar 3},\rep{1})$,
$(\rep{1},\rep{3})$, $(\rep{1},\rep{\bar 3})$ are projected out. In
terms of the representations in the diamond in Table \ref{diamond}, 
this amounts to keeping only the
elements in the horizontal and vertical axes.
This is the analogue of projecting out all triplets in the case of a
single $SU(3)$ structure as we did in ref.\ \cite{GLW}. In that case 
it also removed
all $O(6)$ vectors (or equivalently all one-forms) from the spectrum. 
For a generalized 
\stt\ structure we are lead to project out the vectors of $O(6,6)$, 
which  decompose under 
\stt\  precisely as $\rep{12} \to (\rep{3},\rep{1}) \oplus (\rep{\bar
3},\rep{1}) \oplus(\rep{1},\rep{3}) \oplus (\rep{1},\rep{\bar 3})$.
Note that projecting out $O(6,6)$ vectors does not imply
projecting out all $O(6)$ vectors. For a generic \stt\ structure, 
there are $O(6)$ vectors
(or equivalently one forms) that remain in the spectrum, as for example those
contained in $U_{ \bf 1, \bf 1}$. 
Whenever the structure is not a single SU(3), 
this representation, which is not
projected out, contains a one-form, and the same is true
for all the other representations in the horizontal axis of the diamond.

After this projection both type II theories have two gravitinos and
two Weyl fermions (dilatinos) in the 
$(\rep{1},\rep{1})$ representations. They reside in the gravitational
multiplet and the `universal' tensor multiplet respectively.
Furthermore, eq.\ \eqref{decom88} shows that there is a pair of Weyl
fermions in the representations 
$(\rep{3},\rep{3})\oplus(\rep{\bar 3},\rep{\bar 3})$ and a pair 
in the $(\rep{\bar 3},\rep{3}) \oplus(\rep{3},\rep{\bar 3})$. These
fermions are members of vector- or hypermultiplets depending on which
type II theory is being considered.

The bosonic fields in the NS sector can be similarly decomposed under
\stt. It is convenient to use the combination 
$E_{MN} = g_{MN} + B_{MN}$ 
of the metric and the B-field since from a string theoretical point it
is a tensor product of a left and a right NS-mode excitation.
As a consequence it decomposes under \stt\ as
\begin{equation}
\begin{aligned}
E_{\mu \nu}&: (\rep{1},\rep{1})_{\pm\rep{2}} \oplus
(\rep{1},\rep{1})_{\bf{T}}\ ,\\
 E_{\mu m}&: (\rep{1},\rep{3})_{\pm\rep{1}} \oplus (\rep{1},\rep{\bar 3})_{\pm\rep{1}}\ , \\
E_{m \mu}&: (\rep{3},\rep{1})_{\pm\rep{1}} \oplus (\rep{\bar 3},\rep{1})_{\pm\rep{1}}\ ,  \\
 E_{mn}&: (\rep{3},\rep{3})_{\rep{0}} \oplus (\rep{\bar 3},\rep{\bar
3})_{\rep{0}} \oplus (\rep{\bar 3},\rep{3})_{\rep{0}}
\oplus(\rep{3},\rep{\bar 3})_{\rep{0}} \ ,
\end{aligned}
 \end{equation}
 where ${\bf T}$ denotes the antisymmetric tensor. 
Projecting out 
the  representations $(\rep{3},\rep{1})$, $(\rep{\bar 3},\rep{1})$,
$(\rep{1},\rep{3})$, $(\rep{1},\rep{\bar 3})$ leaves only 
$E_{\mu \nu}$ and $E_{mn}$ in the spectrum. From a four-dimensional point
of view $E_{\mu \nu}$ corresponds to the graviton and an antisymmetric
tensor while $E_{mn}$ represent scalar fields. The latter can be viewed as
paramterizing the deformations of the \stt\ structure or equivalently 
as deformations of  the pure spinors $\Phi^\pm$. More precisely,
keeping the normalization of the pure spinors fixed, $\delta \Phi^+$ transforms
in the $(\rep{\bar 3},\rep{ 3})$, while $\delta \Phi^-$ transforms in the 
$(\rep{\bar 3},\rep{ \bar 3})$ (and $\delta \bar \Phi^+$, $\delta \bar \Phi^-$
transform in the  complex conjugate representations, $(\rep{ 3},\rep{\bar 3})$
and  $(\rep{ 3},\rep{ 3})$ respectively).
  
Finally we decompose the fields in the RR-sector.
Here the bosonic fields arise from the decomposition of 
$(\rep{8}_S, \rep{8}_C)$ for type IIA and $(\rep{8}_S, \rep{8}_S)$ for
type IIB. One finds (after projecting out the triplets)
\beq\label{decompRR}
\begin{aligned}
{\rm IIA}: \quad(\rep{8}_S, \rep{8}_C) \to & \ 
(\rep{1},\rep{{ 1}})_{\pm\rep{1},\rep{0}} \oplus 
 (\rep{3},\rep{{ 3}})_{\rep{0}}  \oplus (\rep{\bar 3},\rep{\bar 3})_{\rep{0}}
 \oplus (\rep{3},\rep{{\bar 3}})_{\rep{1}}  \oplus (\rep{\bar
   3},\rep{3})_{\rep{-1}}\ ,
\\
{\rm IIB}: \quad(\rep{8}_S, \rep{8}_S) \to &\
(\rep{1},\rep{{ 1}})_{\pm\rep{1},\rep{0}} \oplus 
 (\rep{3},\rep{{ 3}})_{\rep{1}}  \oplus (\rep{\bar 3},\rep{\bar 3})_{\rep{-1}}
 \oplus (\rep{3},\rep{{\bar 3}})_{\rep{0}}  \oplus (\rep{\bar
   3},\rep{3})_{\rep{0}}\ .
\end{aligned}
\eeq
In type IIA the RR sector contains gauge potentials of odd degree. The
decomposition \eqref{decompRR} naturally groups these into helicity
$\pm 1$ and helicity $0$ states from a four-dimensional point of view.
This leads us to define
\beq\label{CAexp}
\begin{aligned}
\cC_0^- &= A_{(0,1)} +  A_{(0,3)}+  A_{(0,5)} \simeq 
(\rep{1},\rep{1})_{\rep{0}}  \oplus  (\rep{3},\rep{3})_{\rep{0}} \oplus  (\rep{\bar 3},\rep{\bar 3})_{\rep{0}}\ ,\\
\cC_1^+ &=A_{(1,0)} +  A_{(1,2)}+  A_{(1,4)}+A_{(1,6)}\simeq 
(\rep{1},\rep{{1}})_{\pm\rep{1}} \oplus 
 (\rep{3},\rep{{\bar 3}})_{\rep{1}} \ \oplus (\rep{\bar
   3},\rep{3})_{\rep{-1}} ,\\
\end{aligned}
\eeq
where $A_{(p,q)}$ is a `four-dimensional' $p$-form and a
`six-dimensional' $q$-form.\footnote{There is
an ambiguity in the representation of the scalar degrees of freedom
arising in the RR-sector. They can be equally well written as a 
four-dimensional two-form. On the other hand, $A_1^+$ includes both the vector and dual vector degrees of freedom.}
$\cC_0^-$ contains  `four-dimensional' scalar degrees of freedom 
and is a sum of odd `six-dimensional' forms while 
$\cC_1^+$ contains `four-dimensional' vectors and is a sum 
 even `six-dimensional' forms.

In type IIB the situation is exactly reversed. Here we define
\beq\label{CBexp}
\begin{aligned}
\cC_0^+ &= A_{(0,0)} +  A_{(0,2)}+  A_{(0,4)} +A_{(0,6)} \simeq (\rep{1},\rep{{1}})_{\rep{0}} \oplus 
 (\rep{3},\rep{{\bar 3}})_{\rep{0}}  \oplus  (\rep{\bar
   3},\rep{3})_{\rep{0}}\ ,\\
\cC_1^- &=A_{(1,1)} +  A_{(1,3)}+  A_{(1,5)}\simeq (\rep{1},\rep{1})_{\rep{1}}  \oplus  (\rep{3},\rep{3})_{\rep{1}} \oplus  (\rep{\bar 3},\rep{\bar 3})_{\rep{-1}}\ .\\
\end{aligned}
\eeq

As expected all these fields combine into $N=2$ multiplets, 
as shown in Tables~\ref{N=2multipletsA} and \ref{N=2multipletsB}.
\begin{table}[h]
\begin{center}
\begin{tabular}{|c|c|c|} \hline
 \rule[-0.3cm]{0cm}{0.8cm}
multiplet & \stt rep. & bosonic field content\\ \hline
\rule[-0.3cm]{0cm}{0.8cm}
 gravity multiplet&$(\rep 1, \rep{1}) $ & $ g_{\mu \nu}, \cC^{+}_1$ \\ \hline
\rule[-0.3cm]{0cm}{0.8cm} 
{tensor multiplet} & $(\rep 1, \rep 1)$& $ B_{\mu\nu}, \phi, \cC_{0}^{-} $ \\ \hline
\rule[-0.3cm]{0cm}{0.8cm}
 vector multiplets& $(\rep{3}, \rep{\bar 3})$ &  $\cC^+_{1}, \delta \Phi^+
$\\ \hline
\rule[-0.3cm]{0cm}{0.8cm}
{hypermultiplets } & $(\rep{3}, \rep{3})$&   $ \delta \Phi^-, \cC^-_0 $ \\
\hline
\end{tabular}
\caption{\small 
\textit{N=2 multiplets in type IIA}}\label{N=2multipletsA}
\end{center}
\end{table}
\begin{table}[h]
\begin{center}
\begin{tabular}{|c|c|c|} \hline
 \rule[-0.3cm]{0cm}{0.8cm}
multiplet & \stt rep. & bosonic field content\\ \hline
\rule[-0.3cm]{0cm}{0.8cm}
 gravity multiplet&$(\rep 1, \rep{1}) $ & $g_{\mu \nu}, \cC^{-}_1$ \\ \hline
\rule[-0.3cm]{0cm}{0.8cm} 
{tensor multiplet} & $(\rep 1, \rep{1})$& $B_{\mu\nu}, \phi, \cC^{+}_{0} $ \\ \hline
\rule[-0.3cm]{0cm}{0.8cm}
 vector multiplets& $(\rep{3}, \rep{3})$ &  $ \cC^-_{1},
 \delta \Phi^- $\\ \hline
\rule[-0.3cm]{0cm}{0.8cm}
{hypermultiplets } & $(\rep{3}, \rep{\bar 3})$&   $ \delta \Phi^+, \cC^+_0 $ \\
\hline
\end{tabular}
\caption{\small 
\textit{N=2 multiplets in type IIB}}\label{N=2multipletsB}
\end{center}
\end{table}
We see that the fields arrange nicely and (mirror) symmetrically into
multiplets of a given Spin(6,6) chirality. Mirror symmetry amounts to
a exchange of even and odd Spin(6,6) chirality, or to an exchange of
one $\rep{3}$ into a $\rep{\bar 3}$. This is the analogue  
of the exchange between $\rep{6} \oplus \rep{\bar 3}$ and $\rep{8}
\oplus \rep{1}$ proposed in \cite{FMT} for a single $\SU(3)$
structure.  From these tables it should be clear that \stt\ structure
is the relevant one for $N=2$ effective actions coming from type II
theories.  


\section{$N=2$ and $N=1$ superpotentials}
\label{superpotential}


In this section we show that the $\N = 2$ Killing prepotentials and
the $\N =1$ superpotential found for $\SU(3)$ structures in \cite{GLW}
have exactly the same functional form when the structure is
generalized to \stt.   

The $N=2$ analogue of the $N=1$ superpotential and the $N=1$ $D$-term
are encoded in the Killing prepotentials  $\cP^x$, $x=1,2,3$. These, 
together with its derivatives, determine the scalar potential
\cite{N=2review}. The Killing prepotentials can equivalently be
expressed in terms of the $SU(2)$\footnote{The four-dimensional $N=2$
  theory has a local $SU(2)_R$ symmetry which rotates the two
  (complex)  gravitinos $\psi_{A\,\mu}$ into each other. In  ten
  dimensions it arises from the $O(2)$ rotation of the two
  ten-dimensional Majorana-Weyl fermions into each other.}
gravitino mass matrix $S_{AB}$,  via 
\begin{equation}
\label{4dsusylaws}
   S_{AB} =
      \frac{\ii}2\, \ee^{\tfrac12 K_V}\sigma^x_{AB} \cP^x , 
   \qquad
   \sigma_{AB}^x =
      \begin{pmatrix}
         \delta^{x1} - i \delta^{x2}& -\delta^{x3} \\ 
         -\delta^{x3} & - \delta^{x1} - i \delta^{x2} 
      \end{pmatrix} , 
\end{equation}
where $K_V$ is the K\"ahler potential of the vector multiplets.
The gravitino mass matrix $S_{AB}$ is obtained from the supersymmetry
transformation of the four-dimensional $\N=2$ gravitinos, which has
the generic form 
\beq 
\label{defS}
   \delta \psi_{A\,\mu} = D_\mu \varepsilon_A
      + i \gamma_\mu S_{AB} \varepsilon^B \ , \ A=1,2
\eeq
The four dimensional gravitinos $\psi_{A\, \mu}$ are related to the
ten dimensional ones $\Psi_M$ by \cite{GLW} 
\beq
\hat \Psi_{\mu}^A := {\Psi}_{\mu}^A 
      + \tfrac{1}{2} \Gamma_{\mu}{}^m {\Psi}_m^A= 
       \psi_{A \, \mu +} \otimes \eta^A_\pm
   + \psi_{A\,\mu-} \otimes \eta^A_\mp  +\ldots 
\eeq
where no sum over $A$ is taken on the right hand side, and the $\pm$
are correlated to the chirality of the ten-dimensional spinor, that we
take to be negative (positive) for $A=1$ ($2$) in IIA, and negative
for $A=1,2$ in IIB. In this expression, the dots correspond to the
triplets. 

The supersymmetry transformation of the gravitinos for the democratic
formulation \cite{Bergshoeff} in Einstein frame is  
\begin{multline}
\label{susygravitinoIIA}
   \delta \Psi_M = D_M \epsilon 
      - \frac{1}{96}\ee^{-\phi/2}\left(
         \Gx_M\,^{PQR} H_{PQR} - 9 \Gx^{PQ} H_{MPQ}\right)P\epsilon  \\
      - \sum_{n} \frac{\ee^{(5-n)\phi/4}}{64\,n!} 
         \left[ (n-1)\Gx_M{}^{N_1\dots N_{n}} 
            - n(9-n) \delta_M{}^{N_1}\Gx^{N_2\dots N_n} \right]
         F_{N_1...N_{n}}\, P_n\, \epsilon  \, .
\end{multline}
In this expression, $n=0,2,4,6,8$, $P = \Gamma_{11}$ and $P_n =
(\Gamma_{11})^{n/2} \sigma^1$ for IIA. For IIB we have instead a sum
over $n=1,3,5,7,9$, $P = -\sigma^3$ and $P_n = i \sigma^2$ for
$n=1,5,9$ and $P_n = \sigma^1$ for $n=3,7$. 

In order to get $S_{AB}$, we need to  project the  supersymmetry
transformation of the ten-dimensional shifted gravitino $\delta \hat
\Psi_{\mu}$ onto the SU(3)-singlet parts. The relevant  projector for
type IIB is   
\beq \label{proj}
\Pi= \left( \begin{array}{c} \Pi^1_- \\ \Pi^2_- \end{array} \right)=
\left( \begin{array}{c}  \id \otimes \,  \eta^1_-  \bar \eta^{1}_-  \\
      \id \otimes  \, \eta^2_-  \bar \eta^{2}_- \end{array} \right)
\eeq
(we are using $\bar \eta^A_\pm \eta^A_\pm=1$). For type IIA, we have instead
$\Pi^1_-$ and $\Pi^2_+=  \id \otimes \left(\eta^2_+ \otimes \bar
\eta^{2}_+\right)$.  In the following we
show the details of the type IIB calculation but only give the results
for type IIA since it  follows straightforwardly.

Inserting the projector (\ref{proj}) in $\delta \hat \Psi_{\mu}$, we get
\bea
\left( \begin{array}{c} \delta \psi_{\mu\, +}^1  \\ \delta \psi_{\mu\, +}^2 \end{array} \right)&=& \left( \begin{array}{c} D_{\mu} \epsilon^1_+ \\ D_{\mu} \epsilon^2_+ \end{array} \right)
- \frac{1}{2} \left( \begin{array}{c} \gamma_{\mu} \epsilon^1_-  \  \bar \eta^1_- \gamma^m D_m \eta^1_+ \\ \gamma_{\mu} \epsilon^2_- \ 
\bar \eta^2_- \gamma^m D_m \eta^2_+ 
\end{array} \right) 
  + \frac{1}{48} 
\left( \begin{array}{c} \gamma_{\mu} \epsilon^1_- \  H_{pqr} \, \bar \eta^1_- \gamma^{pqr} \eta^1_+ \\ -\gamma_{\mu} \epsilon^2_- \ 
H_{pqr} \, \bar \eta^2_- \gamma^{pqr} \eta^2_+ 
\end{array} \right) \nn \\
&& - \frac18 \left( \begin{array}{c} - \gamma_{\mu} \epsilon^2_-  \, \ee^{\phi} 
\frac{ 1}{n!} F^-_{i_1...i_n}\, \bar \eta^1_- \gamma^{i_1 ... i_n} \eta^2_+ \\ \gamma_{\mu} \epsilon^1_- \ \ee^{\phi}  
\frac{1}{n!} \sigma(F^-)_{i_1...i_n} \, \bar \eta^2_- \gamma^{i_1 ... i_n} \eta^1_+ 
\end{array} \right) \ ,
\eea
where we have written the expressions in terms of string frame metric
$g=\ee^{\phi/2} g_E$. Furthermore
$F^{-}=F_1 + F_3+F_5$ is the sum of odd internal RR field strengths,
and $\sigma(F^-)= -F_1+F_3-F_5$ is the combination of forms that
appears in the Mukai pairing, Eq.(\ref{mukai})
($\sigma(F_\epsilon)=F_\epsilon^{T}$ in the spinor language) . From
this we read off 
\bea \label{Sbil}
 S_{11}&=&  \frac{\ii}{2}  \, \bar \eta^1_- \gamma^m D_m \eta^1_+ -
\frac{\ii}{48} H_{pqr} \, \bar \eta^1_- \gamma^{pqr} \eta^1_+ \nn \ , \\
 S_{22}&=& \frac{\ii}{2} \, \bar \eta^2_- \gamma^m D_m \eta^2_+ + \frac{\ii}{48} H_{pqr} \, \bar \eta^2_- \gamma^{pqr} \eta^2_+ \nn \ , \\
 S_{12}&=& \frac{\ii}{8}  \,\ee^{\phi} \frac{1}{n!}
F^-_{i_1...i_n} \, \bar \eta^1_- \gamma^{i_1 ... i_n} \eta^2_+  \ , 
\nn \\
 S_{21}&=&  \frac{\ii}{8}  \,\ee^{\phi} \frac{1}{n!}
\sigma(F)^-_{i_1...i_n} \, \bar \eta^2_- \gamma^{i_1 ... i_n} \eta^1_+\ .
\eea
Multiplying by a volume form $\epsilon_g$ and using
(\ref{mukaispinors}), we can write these expressions in terms of Mukai
pairings. $S_{12}$ is the simplest one,  
\beq
\begin{aligned}
   S_{12} \, \epsilon_g
      = S_{21} \, \epsilon_g
      &=  \ii \tr \, (\eta^2_+ \bar \eta^1_-  \gamma^{i_1 ... i_n} )
         \frac{1}{n!}\ee^\phi F^-_{i_1...i_n} \epsilon_g 
      = \ee^\phi\tr ( (\Phi^-_0)^t_\epsilon F^-_{\epsilon} )
         \, \epsilon_g \\
      &= - \tfrac18\,\ee^\phi \langle \Phi_0^-, F^- \rangle 
      = - \tfrac18\,\ee^\phi \langle \Phi^-, G^- \rangle \ , 
 \end{aligned}
 \eeq
where $\Phi_0^-$ is defined in (\ref{purespinors}). In the third
equality we have used $(\Phi_0^-)^t_\epsilon = \ii \eta^2_+ \bar
\eta^1_-$ (see Appendix \ref{app:spinors} for more details) and we
recall that $F$ is related to $F_\epsilon$ by
(\ref{chichiepsilon}). In the first equality, we use
$\sigma(F)_\epsilon=F_\epsilon^{T}$. Finally, in the last equality we
have defined the RR flux $G$ through 
\beq
F =  \dd C - H \wedge C= \ee^B G  \ .
\eeq
$G$ is the flux for the potentials $A$ used in the previous section, namely
 \beq\label{GAdef}
G^+ = \dd A_0^- , \qquad 
G^- = \dd A_0^+ .
\eeq
which implies that $A$ is related to $C$ by $C=\ee^B A$. 

The diagonal pieces in $S_{AB}$ require a bit more work. It is easier
to show that they can also be expressed in terms of
Mukai pairings by working backwards, i.e. starting from the latter and arriving
at the bilinears in (\ref{Sbil}).   
 Using the relation~\eqref{Gdecomp} we have
\begin{equation}
\label{dexp}
\begin{aligned}
   \dd \Phi^+_0 &= \tfrac{1}{2} 
         \left[ \gamma^m , D_m (\eta^1_+\bar{\eta}^2_+ ) \right]_+ \\
      &= \tfrac{1}{2} \left[
         (\gamma^mD_m\eta^1_+)\bar{\eta}^2_+
         + (\gamma^m \eta^1_+) (D_m\bar{\eta}^2_+)
         + (D_m \eta^1_+) (\bar{\eta}^2_+\gamma^m)
         + \eta^1_+ (D_m\bar{\eta}^2_+ \gamma^m) \right] . 
\end{aligned}
\end{equation}
Similarly
\begin{equation}
\label{Hexp}
   H \wedge \Phi^+_0 
      = \tfrac{1}{48}H_{mnp}\left[
         \gamma^{mnp}\eta^1_+\bar{\eta}^2_+
         + 3\gamma^{mn}\eta^1_+\bar{\eta}^2_+\gamma^p
         + 3\gamma^m\eta^1_+\bar{\eta}^2_+\gamma^{np}
         + \eta^1_+\bar{\eta}^2_+\gamma^{mnp} \right] .
\end{equation}
Now we have by chirality and the symmetry of the gamma matrices
\begin{equation}
   \bar{\eta}_-\eta_+ 
      = \bar{\eta}_-\gamma^m\eta_+ 
      = \bar{\eta}_+\gamma^m\eta_+ 
      = 0 .
\end{equation}
Hence, we have
\begin{equation}
\begin{aligned}
 \tfrac18  \mukai{\Phi^-}{\dd\Phi^+}
  	&= \tfrac18
     \mukai{\Phi^-_0}{(\dd\Phi^+_0-H\wedge\Phi^+_0)}
      = -  \tr\left[\left(\Phi^-_0\right)^t_\epsilon 
         \left(\dd\Phi^+_0-H\wedge\Phi^+_0\right)_\epsilon
         \right]\epsilon_g \\
      &= -  \left [\tfrac{\ii}{2} \bar{\eta}^1_-\gamma^mD_m\eta^1_+
         - \tfrac{\ii}{48}H_{mnp}\bar{\eta}^1_-\gamma^{mnp}\eta^1_+
         \right]\epsilon_g = - S_{11} \, \epsilon_g,
\end{aligned}
\end{equation}
where only the first terms in~\eqref{dexp} and~\eqref{Hexp} survive. 
Similarly, one shows that
\beq
 \tfrac{1}{8} \mukai{\Phi^-}{\dd \bar \Phi^+} = S_{22}  \, \epsilon_g \,
\eeq
where now the last terms of the expressions (\ref{dexp}) and
(\ref{Hexp}) corresponding to $\bar \Phi_0^+$ are the only ones that
survive when inserted in the Mukai pairing.  

Collecting all the pieces together, we get for the matrix $S_{AB}$ in
type IIB
\begin{align}
\label{SIIB}
   S_{AB}^{(4)}(\text{IIB}) &= 
      \ee^{\frac{1}{2}K^-}
         \left(\begin{array}{cc}\dstyle 
             -\ee^{\frac{1}{2}K^++\phi^{(4)}}
               \revmukai{\dd\pure^+}{\pure^-} & \dstyle
               - \tfrac{1}{2\sqrt2}\,\ee^{2\phi^{(4)}}
               \revmukai{G^-}{\pure^-} 
            \\*[0.5em] \dstyle
             \, - \tfrac{1}{2\sqrt2}\,\ee^{2\phi^{(4)}}
            \revmukai{G^-}{\pure^-} & \dstyle
            \ee^{\frac{1}{2}K^++\phi^{(4)}}
               \revmukai{\dd\bar\pure^+}{\pure^-}
         \end{array}\right) .
\end{align}
In this expression
the superscript $(4)$ indicates that in (\ref{defS}) we are 
using  the natural metric on $T^{1,3}$: $g_{\mu \nu}^{(4)} = \ee^{-2
  \phi^{(4)}} g_{\mu \nu}$. The four dimensional dilaton $\phi^{(4)}$
is related to the ten dimensional one and the string frame metric by
$\phi^{(4)}=\phi-\tfrac14 \ln \det g_{mn}$. The K\"ahler potentials
$K^\pm$ are defined in (\ref{Kahlerpot}) and we have used that all the
six-forms  are related by the normalization condition  
\beq 
\label{volume}
   \epsilon_g = \tfrac{1}{8}\ii\mukai{\Phi^\pm}{\bar \Phi^\pm}
      = \tfrac{1}{8}\ee^{-K^\pm}=\ee^{-2\phi^{(4)}+2\phi} \ .
\eeq
Note that $S_{AB}$ is naturally a section of ($\Lambda^6 F^*)^{-1/2}$.

The calculation for type IIA follows straightforwardly, and gives 
\begin{align}
\label{SIIA}
   S_{AB}^{(4)}(\text{IIA}) &= 
      \ee^{\frac{1}{2}K^+}
         \left(\begin{array}{cc}\dstyle 
             \ee^{\frac{1}{2}K^-+\phi^{(4)}}
               \revmukai{\dd \pure^-}{\pure^+} & \dstyle
             \tfrac{1}{2\sqrt2}\,\ee^{2\phi^{(4)}}
                 \revmukai{G^+}{\pure^+} 
            \\*[0.5em] \dstyle
               \tfrac{1}{2\sqrt2}\,\ee^{2\phi^{(4)}}
                \revmukai{G^+}{\pure^+} & \dstyle
            - \ee^{\frac{1}{2}K^-+\phi^{(4)}}
               \revmukai{\dd \bar \pure^-}{\pure^+}
         \end{array}\right) .
\end{align}

The gravitino mass matrices obtained have excately the same functional form
in terms of $\Phi^\pm$ as the one obtained in \cite{GLW} for a single
SU(3) structure, confirming the claim made there.\footnote{The
  differences in factors are due to different conventions for the
  normalizations of the spinors, while $S_{11}$ and $S_{22}$ in type
  IIA are interchanged with respect to the expressions in \cite{GLW}
  because we have taken opposte conventions for the chiralities of the
  type IIA spinors.}  They are symmetric under the mirror exchange 
$\Phi^+ \leftrightarrow  \Phi^-$, $G^+ \leftrightarrow G^-$. 

Given the $\N=2$ Killing prepotentials, the computation of the  $\N=1$
superpotential is exactly the same as for a single SU(3) structure. We
will therefore not show the details, worked out in 
\cite{GLW}, but just quote the result\footnote{For orientifold
  compactification on \stt\ manifolds the superpotential has been
  computed in \cite{iman} by reducing the ten-dimensional gravitino
  mass term.}  
\begin{equation}
\label{WA}
\begin{aligned}
   \mathcal{W}_{\text{IIA}} &= 
     \cos^2\phione\,\ee^{\ii\phitwo}\langle \pure^+, \dd
     \pure^-\rangle 
      -\sin^2\phione\,\ee^{-\ii\phitwo} \langle \pure^+, \dd \bar
      \pure^-\rangle \\ & \qquad \qquad \qquad 
      + \frac{1}{2\sqrt{2}}\sin2\phione\,\ee^\phi \revmukai{G^+}{\pure^+}  ,
\end{aligned}
\end{equation}
and
\begin{equation}
\begin{aligned}
\label{WB}
   \mathcal{W}_{\text{IIB}} &= 
    -\cos^2\phione\,\ee^{\ii\phitwo}  \revmukai{\dd\pure^+}{\pure^-}
      +\sin^2\phione\,\ee^{-\ii\phitwo}\revmukai{\dd\bar\pure^+}{\pure^-}
      \\ & \qquad \qquad \qquad 
      -  \frac{1}{2\sqrt{2}}\sin2\phione\,\ee^\phi \revmukai{G^-}{\pure^-} .
\end{aligned}
\end{equation}
where $\alpha$ and $\beta$ parameterize the $U(1)_R\in SU(2)_R$ of
$\N=1$, namely the $\N=1$ supersymmetry parameter $\varepsilon$ is
given in terms of the $\N=2$ parameters $\varepsilon_A$ by 
\begin{equation}
   \varepsilon_A = \varepsilon n_A , \qquad
   n_A = \begin{pmatrix}a \\ b \end{pmatrix}\ , \quad
   a=\cos \alpha\, \ee^{-\ii\beta/2} \ , \ \  
   b=\sin \alpha \,\ee^{\ii\beta/2}\ . 
\end{equation}
The difference between the \stt\ and $\SU(3)$
superpotential is in the form of the pure spinors, which leads to the
appearance of new terms involving the five-form $\dd \Phi^+_4$. As we
will see in the next section, these are the mirrors of magnetic fluxes
missing in pure $\SU(3)$ structure constructions.


\section{Mirror of magnetic fluxes}
\label{magnetic}


Thus far we rewrote the ten-dimensional type II supergravity in a
background which admits an \stt\ structure. In this section we
consider an actual compactification so that the background 
$M^{9,1}=M^{3,1}\times M^{6}$ where $M^{6}$ is a compact manifold
with \stt\ structure. Such reductions in the special case of a pure
$\SU(3)$ structure were discussed in ref.~\cite{GLW}. The analysis
here is completely analogous and therefore we only briefly review this
step. In addition, we will truncate the degrees of freedom in the
forms $\Phi^\pm$ to a finite dimensional space, giving a conventional
effective $N=2$ supergravity theory on $M^{3,1}$. In the case of the
Calabi--Yau this truncation translates into keeping only harmonic
forms and describes the moduli of the Calabi--Yau manifold. As we will
see, in general situations, it is more complicated. This is discussed
in section~\ref{Hmirror} as well as the appendix~\ref{app:expand}.    

The generic case will be considered in the next section. In this
section we 
concentrate on a particular subclass of compactifications for which
one obtains the mirror dual of compactifications on Calabi-Yau
manifolds with magnetic $H_3$-flux. This case was missing in
refs.~\cite{GLMW,GLW} and as a consequence the final results were not
mirror symmetric. Here we close this gap and suggest a completely
mirror symmetric background. Related work has been performed in
refs.~\cite{MR,STW,iman} and we comment on the relation in
section~\ref{sec:nongeo}.  

By way of comparison we first briefly consider the case of
compactification on a Calabi--Yau manifold with generic $H_3$-flux in
the language of generalised structures and identify the truncation. We
then discuss the analogous structure for the mirror symmetric
background.  


\subsection{Generalised geometry and $H_3$-flux}
\label{sec:H3}

Let us review the derivation of the low-energy effective action
arising from a compactification on a Calabi--Yau manifold $M_6$ with
general $H_3$-flux~\cite{Michelson,TV,DallAgata,LM,GM,DSV,DFTV,GLW}. 

One starts by identifying the moduli. Since we want to consider
non-trivial $H_3$ flux we first split the (local) potential
$B$ into flux and moduli pieces 
\begin{equation}
   B = \Bfl + \Bdel, \qquad \qquad
   \dd\Bfl = H_3, \quad \dd\Bdel = 0 .
\end{equation}
The usual Calabi--Yau moduli correspond to expanding the K\"ahler form
$J$, the modulus part $\Bdel$ and the holomorphic three-form $\Omega$
on $M_6$ in terms of forms which are harmonic with respect to the
metric defined by the $\SU(3)$ structure $(J,\Omega)$.  

Specially one expands the three-form $\Omega$ in terms of a symplectic
basis of harmonic three-forms  
\begin{equation}
   \alpha^{(0)}_I, \beta^{(0)I} \in H^3(M_6,\bbR) \ ,
      \qquad I=0,\dots,h_{2,1},
\end{equation}
with
\begin{equation}
\label{o-sympl}
   \int_{M_6} \mukai{\alpha^{(0)}_I}{\beta^{(0)J}} = \delta_I{}^J ,
\end{equation}
and all other pairings vanishing, where we have written the symplectic
structure in terms of the Mukai pairing $\mukai{\cdot}{\cdot}$. One
similarly introduces a set of even harmonic forms to expand $J$ and
$\Bdel$:   
\begin{equation}
\begin{aligned}
   \omega^{(0)}_0 = 1 &\in H^0(M_6,\bbR), &\qquad
   \omega^{(0)}_a &\in H^2(M_6,\bbR), \\
   \tilde{\omega}^{(0)0} &\in H^6(M_6,\bbR), &\qquad
   \tilde{\omega}^{(0)a} &\in H^4(M_6,\bbR), 
\end{aligned}
\end{equation}
with $a=1,\dots,h_{1,1}$ and
\begin{equation}
\label{ab-sympl}
   \int_{M_6} \mukai{\omega^{(0)}_A}{\tilde{\omega}^{(0)B}} 
      = \delta_A{}^B \ , \qquad A,B=0,\dots,h_{1,1} ,
\end{equation}
and all other pairings vanishing. Explicitly, the complex K\"ahler
form is expanded as $\Bdel+\ii J=t^a\omega^{(0)}_a$. Note that the
condition $J\wedge\Omega=0$ implies that  
\begin{equation}
\label{form-conds}
   \omega^{(0)}_a \wedge \alpha^{(0)}_A 
      = \omega^{(0)}_a \wedge \beta^{(0)A} = 0 
      \qquad \forall a, A .
\end{equation}
which is satisfied identically for harmonic forms. 

It is a standard result that there are natural local special
K\"ahler metrics on the moduli spaces of $B+\ii J$ and $\Omega$. These
describe the kinetic energy terms of the moduli in the effective
four-dimensional $N=2$ theory. The properties of special K\"ahler
metrics are discussed in appendix~\ref{app:expand}. In general they
are determined by a holomorphic prepotential $F$. In the Calabi--Yau
context, for the K\"ahler moduli, introducing homogeneous complex
coordinates $X^0=c$ and $X^a=-ct^a$ the corresponding pure spinor can
be written as 
\begin{equation}
   \ee^{-\Bdel}\Phi^+_0 = c\,\ee^{-\Bdel-\ii J} 
      = X^A \omega^{(0)}_A - F_A \tilde{\omega}^{(0)A} ,
\end{equation}
where $F_A=\del F/\del X^A$. Similarly, one has homogeneous complex
coordinates $Z^I$ for the complex structure moduli such that the pure
spinor corresponding to $\Omega$ has the form  
\begin{equation}
   \ee^{-\Bdel}\Phi^-_0 = - \ii\,\Omega 
      = Z^I\alpha^{(0)}_I - \mathcal{F}_I\beta^{(0)I} ,
\end{equation}
where  again $\mathcal{F}_I=\del \mathcal{F}/\del Z^I$.
Using~\eqref{form-conds} one notes that
$\ee^{-\Bdel}\Phi^-_0=\Phi^-_0$. The corresponding K\"ahler potentials
are given by 
\beq\label{Kpm}
\begin{aligned}
   \ee^{-K^+} &= \ii\int_{M^6}\mukai{\Phi_0^+}{\bar\Phi_0^+}
      =  \tfrac{4}{3}c^2\int_{M^6} J \wedge J \wedge J
      = \ii\left(\bar X^A F_A - X^A \bar F_A \right)\ ,\\
   \ee^{-K^-} &=  \ii \int_{M^6}\mukai{\Phi_0^-}{\bar\Phi_0^-}
      = \ii \int_{M^6}\Omega\wedge\bar{\Omega}
      = \ii \left(\bar Z^I \cF_I-Z^I \bar\cF_I \right)\ .
\end{aligned}
\eeq

In deriving the low-energy effective action we assume that the flux
$H_3$ also satisfied the Bianchi identity and equations of motion, and
hence is also harmonic. This means
\begin{equation}
   H_3 = \dd\Bfl = -\p^I \alpha^{(0)}_I + e_I \beta^{(0)I}
\end{equation}
where $\p^I$ are the ``magnetic'' fluxes and $e_I$ the ``electric''
fluxes. Note that for a consistent string theory background the
charges $\p^I$ and $e_I$ must be integral. 

Now in the general expressions for the superpotentials given
section~\ref{superpotential} the pure spinors $\Phi^\pm$ were twisted
by the full potential $B=\Bfl+\Bdel$. It is then natural to introduce
a twisted basis of forms. We write 
\begin{equation}
\begin{aligned}
   \Phi^+ &= \ee^{-B}\Phi^+_0 = X^A \omega_A - F_A \tilde{\omega}^A , \\
   \Phi^- &= \ee^{-B}\Phi^-_0 = Z^I \alpha_I - \mathcal{F}_I\beta^I ,
\end{aligned}
\end{equation}
where the twisted basis forms are given by 
\begin{equation}
\label{twistedbasis}
\begin{aligned} 
   \omega_A &= \ee^{-\Bfl}\omega^{(0)}_A , &\qquad
   \tilde{\omega}^A &= \ee^{-\Bfl}\tilde{\omega}^{(0)A} , \\
   \alpha_I &= \ee^{-\Bfl}\alpha^{(0)}_I , &\qquad
   \beta^I &= \ee^{-\Bfl}\beta^{(0)I} .
\end{aligned}
\end{equation}
Note that $(\omega_A,\tilde{\omega}^A)$ and $(\alpha_I,\beta^I)$ are
no longer of pure degree. Since the Mukai pairing is invariant under
$O(6,6)$ transformations we still have the symplectic structure 
\begin{equation} 
\label{normbasis}
   \int_{M_6} \mukai{\omega_A}{\tilde{\omega}^B} = \delta_A{}^B , \qquad
   \int_{M_6} \mukai{\alpha_I}{\beta^J} = \delta_I{}^J ,
\end{equation}
with the other pairings vanishing. The K\"ahler potentials
$K^\pm=-\ln\ii\int_{M^6}\mukai{\Phi^\pm}{\bar{\Phi}^\pm}$ are
similarly still given by~\eqref{Kpm}. Note that this twisted basis is
an example of a generic truncation, satisfying the necessary
conditions discussed in appendix~\ref{app:expand}.

Crucially the new basis forms are no longer closed. Using the
conditions~\eqref{form-conds}, we find that the only non-zero terms are 
\begin{equation}
\label{nonzero}
\begin{aligned}
   \dd\omega_0 &= - \ee^{-\Bfl}H_3\wedge\omega_0^{(0)} 
      = \ee^{-\Bfl}(\p^I\alpha_I^{(0)}-e_I\beta^{(0)I}) , \\
   \dd\alpha_I &= - \ee^{-\Bfl}H_3\wedge\alpha^{(0)}_I
      = \ee^{-\Bfl}(\p^J\alpha_J^{(0)}-e_J\beta^{(0)J})
         \wedge\alpha^{(0)}_I , \\
   \dd\beta^I &= - \ee^{-\Bfl}H_3\wedge\beta^{(0)I}
      = \ee^{-\Bfl}(\p^J\alpha_J^{(0)}-e_J\beta^{(0)J})
         \wedge\beta^{(0)I} .
\end{aligned}
\end{equation}
Let us introduce a notation ``$\sim$'' to denote equality up to terms
which vanish under the symplectic pairing~\eqref{normbasis} with any
basis form. The non-zero terms are then given by  
\begin{equation}
   \dd\omega_0 \sim \p^I \alpha_I - e_I \beta^I , \qquad
   \dd\alpha_I \sim   e_I \tilde{\omega}^0 , \qquad
   \dd\beta^I \sim  \p^I \tilde{\omega}^0 , 
\end{equation}
where we have used~\eqref{ab-sympl}, and where the first expression is
actually an equality. 

The corresponding low-energy effective action of Calabi-Yau
compactifications with electric and magnetic fluxes has been derived
in refs.~\cite{Michelson,TV,DallAgata,LM,GM,DSV,DFTV,GLW} and for later
reference we recall the Killing prepotentials given in \cite{GLW}
here computed using the expressions above. For type IIA one has
\begin{equation}\label{PdefH}
\begin{aligned}
\cP^1 - \ii\cP^2 
   &= - 2\ii\, \ee^{\frac12 K^- +\phi^{(4)}}  \int_{M^6} 
      \mukai{\Phi^+}{\dd\Phi^-}
   = - 2\ii \ee^{\frac12 K_- +\phi^{(4)}} X^0 
      \big(Z^I e_I - \cF_I m^I \big)\ ,\\
\cP^3 &= \tfrac{1}{\sqrt{2}}\ii\,\ee^{2\phi^{(4)}} 
      \int_{M^6} \mukai{\Phi^+}{G^+}
   =  \tfrac{1}{\sqrt{2}}\ii\,\ee^{2\phi^{(4)}}  
      X^0 \big(\xi^I e_{I}+\tilde\xi_Im^I\big)\ ,
\end{aligned}
\end{equation}
where $\xi^I$, $\tilde{\xi}_I)$ are the RR scalars of type IIA (as
in~\eqref{Cexp} below). 
In type IIB one finds instead 
\begin{equation}\label{PdefIIBH}
\begin{aligned}
\cP^1 - \ii\cP^2 
   &= 2\ii\,\ee^{\frac12 K_+ +\phi^{(4)}}  \int_{M^6} 
      \mukai{\Phi^-}{\dd\Phi^+}
   =  -2\ii\,\ee^{\frac12 K_+ +\phi^{(4)}} 
      X^0 \big(Z^I e_I -\cF_I m^I\big)\ ,\\
\cP^3 &= -\tfrac{1}{\sqrt{2}}\ii\,\ee^{2\phi^{(4)}}
      \int_{M^6} \mukai{\Phi^-}{G^-}
   =  \tfrac{1}{\sqrt{2}}\ii\,\ee^{2\phi^{(4)}}  
      \xi^0 \big(Z^I e_{I} - \cF_I m^I\big)\ ,
\end{aligned}
\end{equation}
where $\xi^0$ is the RR scalar of type IIB. 

To summarize, we have reformulated the moduli and flux expansion in
the conventional Calabi--Yau compactification in terms of a slightly
modified set of twisted forms which naturally include the $H_3$-flux
and are appropriate to the generalised geometry. A key point is that
the elements of the new bases are neither of pure degree nor are
closed. As we will see in the next section, this provides a very
natural ansatz for the corresponding expansion for the mirror
geometries.


\subsection{Generalised geometry and the mirror of $H_3$-flux}
\label{Hmirror}

Following the setup of ref.~\cite{GLW} and in analogy with our
reformulation of the Calabi--Yau compactification with $H_3$-flux, we
now look for some basis of forms on $M^6$ in which to expand the
fields of the ten-dimensional background (summarized in
tables~\ref{N=2multipletsA} and~\ref{N=2multipletsB}). It is clear
from the Calabi--Yau discussion that in general the basis forms in
$\Lambda^*TM^*$ need not be of pure degree, nor closed. 

Physically we are keeping only certain modes in the entire tower of
Kaluza-Klein excitations which correspond to the light
modes of the compactification. Obviously to actually identify this
hierarchy of excitations requires a knowledge of the particular
properties of $M_6$. In the following, rather than fix the manifold
and show that there is a sensible set of light modes, we will simply
assume there is such an expansion and discuss its consistency
conditions. (For a further discussion of when such a truncation exists
see~\cite{mkp}.) 
Indeed, if mirror symmetry can be defined for a Calabi--Yau compactification
with $H_3$ flux, then there must
be some dual compactification for which such a 
hierarchical expansion can be identified.

The general truncation consistency conditions are discussed in detail
in appendix~\ref{app:expand}\footnote{The
  conditions in the special case of a generic $\SU(3)$ structure were
  also analysed recently in~\cite{mkp}.}. 
Since $\Phi^\pm$ and $G^\pm$ are sums of either odd or even forms, our
basis should similarly be in terms of odd or even forms. For the
kinetic terms to make sense (and to have the correct multiplet
structure) we better ensure that the special K\"ahler geometry for the
untruncated $\Phi^\pm$ descends to a special K\"ahler geometry for the
finite number of modes we are keeping.

In general we identify two finite-dimensional subspaces
$\vt^\pm\subset\scs{S^\pm(E)}$ and require $\Phi^\pm$ to lie in
$\vt^\pm$. Explicitly we can expand $\Phi^\pm$ in terms of a basis of
forms 
\begin{equation}
\label{deffin}
\begin{aligned}
   \Ae &= \{\omega_A,\tilde{\omega}^B\} , & A = 0, \dots, \beve , \\
   \Ao &= \{\alpha_I,\beta^J\} , & I = 0, \dots, \bodd .
\end{aligned}
\end{equation}
which define a symplectic structure 
\begin{equation}
\label{symplecticbasis}
   \int_{M^6}\langle \omega_A, \tilde\omega^B\rangle =  \delta_A{}^B\ , \qquad
   \int_{M^6}\langle \alpha_I, \beta^J\rangle = \delta_I{}^J\ ,
\end{equation}
with all other pairings vanishing. For there to be a natural local
special K\"ahler structure on $\vt^\pm/\bbC^*$, these bases must
satisfy a number of other conditions given in detail
in appendix~\ref{app:expand}. Ignoring the
compatibility condition~\eqref{comp1} one can then introduce
holomorphic coordinates and prepotentials as before, and expand the
pure spinors $\Phi^\pm$ as follows 
\begin{equation}
\label{Phiexp}
   \Phi^+ = X^A \omega_{A} - F_A\tilde\omega^A\ , 
   \qquad 
   \Phi^- = Z^I \alpha_I - \cF_I \beta^I \ .
\end{equation}

Generically, however, the compatibility condition~\eqref{comp1}
imposes a relation between the moduli. To avoid this, we will assume,
that~\eqref{comp1} is satisfied by each pair of basis forms 
\begin{equation} 
\label{compbasis}
   \mukai{\omega_A}{V\cdot\alpha_I} 
      = \mukai{\omega_A}{V\cdot\beta^I} 
      = \mukai{\tilde{\omega}^A}{V\cdot\alpha_I} 
      = \mukai{\tilde{\omega}^A}{V\cdot\beta^I}
      = 0\ ,
\end{equation}
for all $V=x+\xi\in E$. These are the analogues of the
conditions~\eqref{form-conds} in the Calabi--Yau case and imply that
the expressions~\eqref{Phiexp} are valid without constraining the
moduli. 
In fact~\eqref{compbasis} further implies that there are no triplet
representations under \stt\ in the expansion which has to hold so that
no additional spin-$\frac{3}{2}$ multiplets are in the light
spectrum. To see this, note, first, that a generic $\chi\in S(E)$
contains eight triplet components as indicated in Table~\ref{diamond}.
Similarly, a generic vector $V\in E$ decomposes into four triplets
$(\rep{3},\rep{1})+(\bar{\rep{3}},\rep{1})
+(\rep{1},\rep{3})+(\rep{1},\bar{\rep{3}})$ under \stt.  
Since the Mukai pairing and the pure spinors $\Phi^\pm$ are singlets, 
the condition
\begin{equation}
\label{no-triplet}
   \mukai{\Phi^+}{V\cdot\chi} = \mukai{\Phi^-}{V\cdot\chi} = 0 , 
   \quad \forall V\in E ,
\end{equation}
is equivalent to setting the eight triple components of $\chi$ to
zero. Given the expansion~\eqref{deffin} and using the fact
that $\mukai{\psi}{V\cdot\chi}=-\mukai{\chi}{V\cdot\psi}$, it is easy
to check that~\eqref{no-triplet} is indeed satisfied for every
basis form. 

The truncated K\"ahler potentials are given by the same expressions as
in the Calabi--Yau case~\eqref{Kpm} and read
\beq
\begin{aligned}\label{Kpmgen}
   \ee^{-K^+} &= \ii\int_{M^6}\mukai{\Phi^+}{\bar\Phi^+}
      = \ii\left(\bar X^A F_A - X^A \bar F_A \right)\ ,\\
   \ee^{-K^-} &=  \ii\int_{M^6}\mukai{\Phi^-}{\bar\Phi^-}
      =  \ii\left(\bar Z^I \cF_I-Z^I \bar\cF_I \right)\ .
\end{aligned}
\eeq

For the Ramond-Ramond fields we expand the combinations
$\cC^\pm_0$ and $\cC^\pm_1$ defined in eqs.~\eqref{CBexp}, \eqref{CAexp}
in terms of the symplectic basis~\eqref{symplecticbasis} as follows 
\begin{equation}\label{Cexp}
\begin{aligned}
\cC^+_0 & = \xi^A \omega_A +\tilde\xi_B\tilde\omega^B\ , \qquad
\cC^-_1  = A_1^I\alpha_I + \tilde A_{1 J}\beta^J\ , \\ 
\cC^-_0 & = \xi^I\alpha_I+\tilde\xi_J\beta^J\ , \qquad
\cC^+_{1}  = A_1^A \omega_A + \tilde A_{1 B}\tilde\omega^B\ . \\ 
\end{aligned}
\end{equation}
$\xi^A$ and $\tilde\xi_B$ are scalars and  $A_1^I$ and $\tilde A_{1 J}$ are
vectors in type IIB while  $\xi^I$, $\tilde\xi_J$, $A_1^A$ and $\tilde
A_{1 B}$ are scalars and vectors of type IIA respectively. In the
following it will sometimes be more convenient to dualize the scalars
of $\cC^+_0$ and $\cC^-_0$ to antisymmetric tensors and, when appropriate,
discuss the effective theory in terms of them.\footnote{The reason is
  that the magnetic fluxes or torsion charges generate masses for some
  of the antisymmetric tensors and the discussion becomes a bit more
  involved in terms of scalar degrees of freedom \cite{LM,DSV,dWST}.} 
Thus we define 
\begin{equation}\label{Cdual}
\begin{aligned}
\cC^+_{2} & =  \tilde C_{2}^A \omega_A + C_{2  B}\tilde\omega^B\ , \qquad
\cC^-_{2}  = \tilde C_{2}^I\alpha_I+ C_{2  J}\beta^J\ , \\ 
\end{aligned}
\end{equation}
where from a four-dimensional point of view 
$\cC^+_{2}$ is dual to $\cC^+_{0}$ and $\cC^-_{2}$ is dual to
$\cC^-_{0}$. At the level of the four-dimensional fields 
the duality relates 
\begin{equation}\label{Cdualf}
\begin{aligned}
   \xi^A \leftrightarrow  C_{2  A}\ ,\qquad 
   \tilde\xi_B  \leftrightarrow\tilde C_{2}^B \ ,\qquad
   \xi^I \leftrightarrow  C_{2  I}\ ,\qquad 
   \tilde\xi_J \leftrightarrow \tilde C_{2}^J\ .
\end{aligned}
\end{equation}

The goal of this section is to find the dual of the magnetic
fluxes. We know that mirror symmetry essentially exchanges $\Leven
T^*M_6$ and $\Lodd T^*M_6$. We also showed in the previous section
that the $H_3$ flux is naturally incorporated in the generalised
geometry picture as non-closed basis forms~\eqref{nonzero}. Thus for
the mirror compactification it is natural to take the same
differential conditions~\eqref{nonzero} but with the roles of odd and
even forms reversed:\footnote{Note that $\dd^2=0$ is automatically
  satisfied.}
\begin{equation}
\label{diffcond}
   \dd \alpha_{0}  \sim  \m^A \omega_{A} + e_A\tilde\omega^A\ , \qquad 
   \dd\omega_{A} \sim -e_A \beta^{0} \ , \qquad 
   \dd\tilde\omega^{A} \sim \m^A \beta^{0} \  .
\end{equation}
Note that as before these relations are only up to terms which vanish
under the symplectic pairing~\eqref{symplecticbasis}. 
Here we have singled out two of the basis forms $\alpha_0$ and
$\beta^0$. This is a familiar property of local special K\"ahler
metrics. The point is that the $\Phi^\pm$ are only defined up to
complex rescalings. From eqs.~\eqref{Kpmgen} we see that $\Phi^\pm \to
c^\pm \Phi^\pm$ amounts to a K\"ahler transformation of
$K^\pm$. Therefore it is possible to go to `special coordinates' where
one of the $X^A$ and one of the  $Z^I$, say $X^0$ and  $Z^0$,  is
scaled to one. This arbitrarily singles out one of each of the basis
elements namely $\omega_0$ and $\alpha_0$, and the dual
$\tilde{\omega}^0$ and $\beta^0$. 

For $\m^A=0$ the conditions~\eqref{diffcond} precisely correspond to
the conditions imposed in ref.~\cite{GLMW} with $e_A$ being the mirror
dual of the electric fluxes. Note that in ref.~\cite{GLMW} it was assumed
that all the basis forms were of pure degree and hence $\m^A$ was
necessarily zero. The generalisation here is that we allow the basis
forms to be of mixed degree. The next step is to show that the
$\m^A$ in~\eqref{diffcond} corresponds to  the mirror dual of the
magnetic fluxes. We do not compute the entire effective action but
instead only focus on the mass terms of the antisymmetric tensor, the
covariant derivatives of the scalars and the Killing prepotential. Let
us discuss these in turn. 

The ten-dimensional type IIA action contains terms of the form
$|G_{2p}|^2$ where $G_{2p}= \dd A_{2p-1}$ is the $2p$-form field strength 
of the $(2p-1)$-form gauge potential $A_{2p-1}$. In the compactified
theory the combination 
$\dd \cC_{2}^- + \df\cC_1^-$ appears where now $\dd $ denotes  the exterior
derivative on $M^6$ while $\df$ is the exterior
derivative on $M^{3,1}$. 
Using \eqref{Cexp} and \eqref{diffcond} we find
\begin{equation}\label{Fexp}
\dd \cC_{2}^- +  \df\cC_1^+ = 
\fs_{2}^A\, \omega_A +
\fst_{2 A}\,\tilde\omega^A\ ,
\end{equation}
where 
\begin{equation}
\begin{aligned}
\fs_{2}^A &= \tilde C_{2}^0\, \m^A + \df A^A_{1}\ , \\
\fst_{2  A} &= \tilde C_{2}^0\, e_A + \df \tilde A^A_{1}
\ .
\end{aligned}
\end{equation}
$\fs_{2}^A$ is invariant under the combined gauge
transformations
\begin{equation}
\delta \tilde C_2^0 = \df \Theta_1 \ ,\qquad \delta A_1^A = -\m^A\Theta_1\ ,
\end{equation}
where $\Theta_1$ is a one-form gauge parameter. We see that by an
appropriate gauge choice one linear combination of vectors $A_1^A$
can be removed from 
the spectrum or in other words they become the longitudinal degree of
freedom of a massive $\tilde C_2^0$. Indeed,
repeating the analysis of ref.~\cite{LM,GLMW} one easily shows that 
the effective action contains terms proportional to 
$\fs_{2}\wedge \fs_{2}$ and $\fs_{2}\wedge *\fs_{2}$.
{}From this we conclude that 
for $\m^A \neq 0$ the antisymmetric tensor $\tilde C_{2}^0$  aquires a mass by 
a Stueckelberg mechanism or in other words by `eating' a vector. 
This is precisley what one finds in Calabi-Yau compactifications 
of type IIB with magnetic fluxes as computed in ref.~\cite{GLMW} and 
thus we have a first crucial check that we have succesfully identified
the mirror dual compactification. 

As a second check let us compute the Killing prepotential 
on the finite subspaces $\vt^\pm$.
Using \eqref{Phiexp},  \eqref{Cexp},  \eqref{diffcond} and 
$G^+ =\dd A_0^-$ we obtain from (\ref{SIIA})
\begin{equation}\label{Pdef}
\begin{aligned}
\cP^1 - \ii\cP^2 
   &= -2\ii\, \ee^{\frac12 K^-+\phi^{(4)}}\int_{M^6}\mukai{\Phi^+}{\dd\Phi^-}
   = -2\ii\, \ee^{\frac12 K^-+\phi^{(4)}} 
      Z^0 \big(X^A e_A + F_A \m^A\big)\ ,\\
\cP^3 &= \tfrac{1}{\sqrt{2}}\ii\,\ee^{2\phi^{(4)}} 
     \int_{M^6}\mukai{\Phi^+}{G^+}
   = \tfrac{1}{\sqrt{2}}\ii\,\ee^{2\phi^{(4)}} 
      \xi^0 \big(X^A e_{A}+F_A\m^A\big)\ .
\end{aligned}
\end{equation}
These are precisely the correct Killing prepotential for the mirror 
dual compactification as can be seen by comparing with eq.~\eqref{PdefIIBH}.
Under the exchange
$X^A \leftrightarrow Z^I,\ F_A~\leftrightarrow~\cF_I$,
$e_{I}~\leftrightarrow~e_{A}, \
\p^{I}~\leftrightarrow~ -\m^A$ the expressions are
identical. 

For completness let us also display the results for type IIB compactifications.
In this case no antisymmetric tensor becomes massive and thus it is
more convenient to 
use the scalars in $\cC^{+}_0$ of \eqref{Cexp} in our discussion. From the
ten-dimensional type IIB action one obtains the combination 
$\dd\cC_{1}^+ + \df\cC^{+}_0$ in the four-dimensional effective action.
Using \eqref{Cexp} and \eqref{diffcond} we find
\begin{equation}\label{covder}
\dd\cC_{1}^- + \df\cC_{0}^+ = 
D\xi^A\, \omega_A +
D \tilde\xi_A\,\tilde\omega^A
\end{equation}
where 
\begin{equation}
D\xi^A = \df\xi^A - \m^AA_\mu^0\ , \qquad
D \tilde\xi_A =\df\tilde\xi_A-e_AA_\mu^0\ .
\end{equation}
We see that, depending on the choice of $\m^A, e_A$, a linear
combination of $\xi^A, \tilde\xi_A$ becomes the longitudinal degree of
freedom of a massive vector $A_{\mu}^0$. Again,
this is precisley what one finds in Calabi-Yau compactifications 
of type IIA with electric and magnetic fluxes as computed in ref.~\cite{LM}.
The corresponding Killing prepotentials are given by 
\begin{equation}\label{PdefIIB}
\begin{aligned}
\cP^1 -\ii\cP^2 
   &= 2\ii\,\ee^{\frac12K^++\phi^{(4)}}
      \int_{M^6}\mukai{\Phi^-}{\dd\Phi^+}
   =  - 2\ii\,\ee^{\frac12 K^+ +\phi^{(4)}} 
      Z^0\big(X^Ae_A +F_A \m^A\big)\ ,\\
\cP^3 &=- \tfrac{1}{\sqrt{2}}\ii\,\ee^{2\phi^{(4)}} 
      \int_{M^6}\mukai{\Phi^-}{G^-}
   =  \tfrac{1}{\sqrt{2}}\ii\,\ee^{2\phi^{(4)}}  
      Z^0\big(\xi^A e_{A}-\tilde\xi_A\m^A\big)\ ,
\end{aligned}
\end{equation}
which again are perfectly mirror symmetric to \eqref{PdefH}.

Let us summarize. By considering compactifications of type IIA 
on a specific class of manifolds with \stt\ we were able to identify
mirror duals of type IIB compactifications on Calabi-Yau threefolds
with generic background $H_3$-flux. The dual manifolds are characterized 
by the condition \eqref{diffcond} which generalize the half-flat conditions 
of ref.~\cite{GLMW}. The new ingredient is a non-zero parameter $\m^A$
which plays the role of a dual magnetic flux.\footnote{This dual 
 background has also been confirmed by identifying mirror symmetric 
$N=1$ domain wall solutions \cite{LV}.} Note that the quantization of
the dual $H_3$ implies that $e_A$ and $\m^A$ are similarly
integral. In simple examples, these conditions are necessarily
satisfied since $e_A$ and $\m^A$ are related to topological
invariants of the manifold.

Instead of giving \eqref{diffcond} we can equally well specify 
differential constraints of $\Phi^\pm$. Using  \eqref{Phiexp} one
obtains 
\begin{equation}
\label{dPhi}
   \dd\Phi^+ = -(X^A e_A +F_Ap^A)\, \beta^0\ ,\qquad 
   \dd\Phi^- = Z^0( \m^A \omega_{A} + e_A\tilde\omega^A) \ .
\end{equation}
This compares with 
\begin{equation}
\label{dPhi-H}
   \dd\Phi^+ = X^0(\p^I \alpha_I - e_I\beta^I) \ ,\qquad 
   \dd\Phi^- = (Z^I e_I - \cF_I \p^I)\, \tilde{\omega}^0\ ,
\end{equation}
for the case of a Calabi--Yau compactification with $H_3$ flux. As
expected we see that mirror symmetry is just exchanging odd and even
forms. Note that the right hand side of $\dd\Phi^-$ in~\eqref{dPhi} is
real and thus we have  
\begin{equation}\label{ghf}
   \dd\im\Phi^- = 0\ .
\end{equation}
The same constraint holds for half-flat manifolds but in that case also 
$J\wedge J$ is closed. Here, this second constraint no longer
holds. Furthermore, since $\alpha_I$ and $\beta^I$ are generically of
mixed degree, $\Phi^-$ is no longer purely a three-form.


\section{Generic \stt\ compactifications}
\label{general}


In the previous section we considered manifolds with \stt\ structure
which can serve as mirror dual compactifications of Calabi-Yau
backgrounds with generic NS-flux. In this section we consider a more
general class of compactifications by relaxing~\eqref{diffcond}
and~\eqref{dPhi}. As before we consider a generic
truncation~\eqref{deffin}, with the triplets projected out, but now
allow for the most general differential conditions which can be
imposed on the two symplectic basis. They read
\begin{equation}\label{diffcond2}
\begin{aligned}
\dd \alpha_{I} & \sim \m^A_I \omega_{A} + e_{IA}\tilde\omega^A\ , & \qquad
\dd\beta^I &\sim q^{IA} \omega_{A} + \p_{A}^I\tilde\omega^A , \\
\dd\omega_{A} &\sim \p_A^I \alpha_I - e_{IA}\beta^{I} \ , & \qquad 
\dd\tilde\omega^{A} &\sim -q^{IA}\alpha_I + \m^A_I \beta^{I} \  ,
\end{aligned}
\end{equation}
where $\m^A_I, e_{IA},q^{IA}, \p_{A}^I$ are four 
$(\beve +1) \times (\bodd +1)$-dimensional constant matrices.
Following the discussion of the previous section, we expect these
matrices to take integer values. 
In order to make the symplectic structure manifest let us introduce a
notation for the two symplectic basis
\beq\label{Xiintr}
\Ae := \left( \begin{array}{c} \omega_A \\ \tilde \omega^B \end{array}
\right) \ , \ \qquad 
\Ao :=\left( \begin{array}{c} \alpha_I \\  \beta^J \end{array} \right)
\ .
\eeq
In terms of $\Ae$ and $\Ao$ eq.~\eqref{diffcond2}
turns into
\beq \label{diffQ}
\dd \Ao \sim \cQ  \, \Ae \ , \qquad \dd \Ae \sim \SSe \cQ^T (\SSo)^{-1} \,  \Ao
\eeq
where
\beq \label{chargem}
{\cal Q}=\left( \begin{array}{cc}  p_I{}^A &  e_{IB} \\ q^{JA} & m^J{}_B
\end{array} \right) \ ,
\eeq 
and $\SSe$ and $\SSo$ are the symplectic structures on $\vt^+$ and
$\vt^-$. 
Note that $\dd \Ao$ and $\dd \Ae$ have to depend on the same matrix 
${\cal Q}$ in order to ensure consistency of 
$\int_{M^6}\mukai{\Ae}{\dd \Ao} =  \int_{M^6}\mukai{\dd\Ae}{\Ao}$.
Furthermore $\dd^2=0$ implies two additional quadratic constraints
\beq
\label{lagQ}
{\cal Q}\SSe{\cal Q}^T\ =\ 0\ =\ {\cal Q}^T (\SSo)^{-1}{\cal Q}\ ,
\eeq 
or explicitly 
\begin{equation}
\label{diffconstraints}
\begin{aligned}
q^{IA} \p_A^J - \p^I_Aq^{AJ} &= 0 \ , &\qquad
\m_I^Ae_{AJ}-e_{IA}\m^A_J &= 0 \ , &\qquad
\m_I^A\p_A^J-e_{IA}q^{AJ} &= 0\ ,\\
q^{AI}\m_I^B - \m_I^Aq^{IB} &= 0  \ , &\qquad
\p^I_Ae_{IB} - e_{AI}\p^I_B &=0 \ , &\qquad
\p^I_A\m^B_I -e_{AI}q^{IB} &= 0 \ .
\end{aligned}
\end{equation}
The `doubly symplectic' charge matrix ${\cal Q}$ has also been
discussed in refs.~\cite{Berglund,Dallaattractor}. 

Note that we can count the number of independent charges
in~$\mathcal{Q}$ as follows. Formally $\mathcal{Q}$ is a linear map
$\mathcal{Q}:\vt^-\to\vt^+$, or equivalently
$\mathcal{Q}\in(\vt^-)^*\otimes\vt^+$. The conditions~\eqref{lagQ}
imply that images of $\mathcal{Q}$ and $\mathcal{Q}^T$ are isotropic 
subspaces, denoted by $L^+:=\img\mathcal{Q}\subset\vt^+$ and
$\bar{L}^-:=\img\mathcal{Q}^T\subset(\vt^-)^*$
respectively. Equivalently, $\mathcal{Q}\in\bar{L}^-\otimes L^+$,
with, as for any linear map, $p:=\dim L^+=\dim \bar{L}^-$. Since $L^+$
and $\bar{L}^-$ are isotropic we have $p\leq\beve+1$ and $p\leq\bodd+1$. 
Furthermore, a $p$-dimensional isotropic subspace in a
$2d$-dimensional symplectic space is determined by $2dp-\frac12p(p-1)$
parameters. Thus counting first the parameters in choosing $L^+$ and
$\bar{L}^-$ and then the $p^2$ independent elements of $\mathcal{Q}$
given $L^+$ and $\bar{L}^-$, we find that generically 
\begin{equation}
\label{dimQ}
   \dim \mathcal{Q} = \begin{cases}
      (2\bodd+3)(\beve+1) & 
         \quad \text{if $\beve \leq \bodd$} \\
      (2\beve+3)(\bodd+1) & 
         \quad \text{if $\bodd \leq \beve$}
   \end{cases}
\end{equation}
corresponding to $p=\beve+1$ and $p=\bodd+1$ respectively. 

The next step is to compute again the Killing prepotentials.
In the type IIA low energy effective action the quantity
$\dd\cC_{2}^- + \df\cC_1^+$ appears exactly as in the previous section
and it again obeys the expansion \eqref{Fexp}. However  due to
\eqref{diffcond2} the coefficients of this expansion now read
\begin{equation}\label{IIA}
\begin{aligned}
\fs_{2}^A & = 
\tilde C_{2}^I \m^A_I+  C_{2  I} q^{AI} + \dd_4 A_1^A\ , \\
\fst_{2  A}& =
\tilde C_{2}^I e_{AI} +  C_{2  I} \p^{I}_A+\dd_4 \tilde A_1^A\ .
\end{aligned}
\end{equation}
Recall that $\dim(\img\mathcal{Q})=p$ with $p\leq\beve+1$ and
$p\leq\bodd+1$. Hence the number of linearly independent massive
antisymmetric tensors $\fs_2^A$ and $\fst_{2A}$ in~\eqref{IIA} is
$p$. Thus if $\beve\geq\bodd$ at most $\bodd+1$ tensors are
massive, and if $\bodd\geq\beve$ at most $\beve+1$ tensors are
massive.  

The Killing prepotentials are always expressed in terms of the
scalar fields.
They can be computed exactly as in the previous
section but now using \eqref{diffcond2} instead of \eqref{diffcond}.
This yields
\begin{equation}\label{Pdef2}
\begin{aligned}
\cP^1 - \ii\cP^2 &= 
-2\ii\,\ee^{\frac12 K^- +\phi^{(4)}}\int_{M^6} \mukai{\Phi^+}{\dd\Phi^-}
\\
&=2\ii\, \ee^{\frac12 K^- +\phi^{(4)}}  \big(V^{-T} \SSo {\cal Q}
V^+
\big)\\
&= 2\ii\, \ee^{\frac12 K^- +\phi^{(4)}} 
\big( -X^A e_{AI} Z^I + X^A \p^{I}_A\cF_I-F_A\m^A_I Z^I 
+ F_A q^{AI}\cF_I   \big)\ ,
\end{aligned}
\end{equation}
and 
\begin{equation}\label{P3def2}
\begin{aligned}
\cP^3 &= \tfrac{1}{\sqrt{2}}\ii\,\ee^{2\phi^{(4)}} 
   \int_{M^6} \mukai{\Phi^+}{G^+} \\
&= \tfrac{1}{\sqrt{2}}\ii\,\ee^{2\phi^{(4)}} 
\big(V_{\xi}^{-T} \SSo {\cal Q}
V^+ + V_{RR}^{+T} \SSe V^+\big)\\
& = \tfrac{1}{\sqrt{2}}\ii\,\ee^{2\phi^{(4)}} 
\left[(X^A(\tilde G_{RR\,A} +e_{AI}\xi^I 
  +\p^I_A\tilde\xi_I)
+F_A(G_{RR}^A + \m_{I}^A \xi^I +q^{AI}\tilde\xi_I)\right]\ ,
\end{aligned}
\end{equation}
where we introduced  the symplectic sections
\begin{equation}\label{Vdef}
\begin{aligned}
V^+
= \left( \begin{array}{c}F_A  \\  X^B\end{array} \right)\ ,
\qquad
V^-= \left( \begin{array}{c}  \cF_I \\ Z^J\end{array} \right)
\ ,\qquad
V^-_{\xi}= \left( \begin{array}{c}  \tilde\xi_I \\ -\xi^J\end{array} \right)
\ ,\qquad
V_{RR}^+=\left( \begin{array}{c}\tilde G_{RR\,A}\\ -G_{RR}^B\end{array}
  \right)\ ,
\end{aligned}
\end{equation}
and expanded
\begin{equation}
G^+ = G_{RR}^A\omega_A + \tilde G_{RR\, A}\tilde\omega^A + \dd A_0^-\ .
\end{equation}
Here $G_{RR}^A, \tilde G_{RR\, A}$ denote the RR-fluxes.\footnote{Note
  that combinations of scalars $(\xi^I,\tilde{\xi}_I)$ which is dual
  to the massive tensors given by~\eqref{IIA} precisely drops out of
  the expression for $\cP^3$ as is required for
  consistency. Alternatively one can formulate the supergravity in a
  redundant form where both scalar degrees of freedom together with
  antisymmetric tensors are kept~\cite{dWST}.}  
Note that $\cP^1 +i\cP^2$ has the same form as the  superpotential
introduced in ref.~\cite{Berglund} where it was inferred from F-theory
considerations. It would be interesting to make the correspondence
with the results of  ref.~\cite{Berglund} more precise.

In the large volume limit the holomorphic prepotential $F$ is a cubic 
function of the scalar fields in the vector multiplets. 
From \eqref{Pdef2} we see that the
matrices $\m_I^A$ and $q^{AI}$ multiply quadratic and cubic terms
while $e_{AI}$ and $\p_A^I$ multiply constant and linear terms. 
{}Mirror symmetry implies that there is a limit where $\cF$ has a
similar expansion. In the next section we discuss the specific example
of flux backgrounds on twisted toroidal compactification in more
detail, hence establishing the relation of these results with those of
ref.~\cite{STW}. 

Let us turn to type IIB. 
In order to see massive tensors occuring one considers the quantity
$\dd\cC_{2}^+ + \df C_1^-$ instead of 
$\dd\cC_{1}^- + \df C_0^+$ as done in \eqref{covder}.
Using \eqref{diffcond2} and \eqref{Cdual} one finds
\begin{equation}
\dd\cC_{2}^+ + \df A_1^- = 
\fs_{2}^I\, \alpha_I +
\fst_{2  J}\,\beta^J \ ,
\end{equation}
where
\begin{equation}\label{IIB}
\begin{aligned}
\fs_{2}^I & = 
- \tilde C_{2}^A \p_A^I+ C_{2  A} q^{AI} + \dd_4 A_1^I\ , \\
\fst_{2  I}& =
 \tilde C_{2}^A e_{AI} -  C_{2  A} \m^A_I+ \dd_4 \tilde A_1^I\ .
\end{aligned}
\end{equation}

The Killing prepotentials are again expressed in terms of scalar
fields. Repeating the calculation of the last section with
\eqref{diffcond} replaced by \eqref{diffcond2} one finds
\begin{equation}\label{PdefIIB2}
\begin{aligned}
\cP^1 -\ii\cP^2 &= 2\ii \ee^{\frac12 K^+ +\phi^{(4)}}  \int_{M^6}
  \mukai{\Phi^-}{\dd\Phi^+} \\
&= 2\ii\, \ee^{\frac12 K^+ +\phi^{(4)}}  \big(V^{-T} \SSo {\cal Q}
V^+\big)
\\
&= 2\ii\, \ee^{\frac12 K^+ +\phi^{(4)}} 
\big( - Z^I e_{IA}X^A - Z^I\m^A_I F_A + \cF_I \p^{I}_A X^A +\cF_I
q^{IA}  F_A  \big)\ ,
\end{aligned}
\end{equation}
and 
\begin{equation}\label{P3defIIB2}
\begin{aligned}
\cP^3 &= -\tfrac{1}{\sqrt{2}}\ii\,\ee^{2\phi^{(4)}} 
   \int_{M^6} \mukai{\Phi^-}{G^-} \\
&= -\tfrac{1}{\sqrt{2}}\ii\,\ee^{2\phi^{(4)}} \big(V^{-T} \SSo {\cal Q}
V_{\xi}^+ 
+ V_{RR}^{-T} \SSo V^- \big)\\
&= -\tfrac{1}{\sqrt{2}}\ii\,\ee^{2\phi^{(4)}}  \left[
Z^I(\tilde G_{RR\,I} -e_{IB}\xi^B
  +\m^A_I\tilde\xi_A)
+\cF_I(G_{RR}^I + \p_{B}^I \xi^B -q^{IA}\tilde\xi_A)
\right]\ ,
\end{aligned}
\end{equation}
where 
\begin{equation}
G^- = G_{RR}^I\alpha_I + \tilde G_{RR\, J}\beta^J + \dd A_0^+\ ,\qquad
V_{\xi}^+= \left( \begin{array}{c}  \tilde\xi_A \\ -\xi^B\end{array} \right)
\ ,\qquad
V_{RR}^-=\left( \begin{array}{c}\tilde G_{RR\,I}\\ -G_{RR}^J\end{array}
  \right)\ ,
\end{equation}
and $G_{RR}^I, \tilde G_{RR\, I}$ again denote the RR-fluxes.

Let summarize the role the different $\mathcal{Q}$-charges take in the
low energy effective theory. Generically 
they always give a mass to some of the light modes. Depending on which
charge is under consideration in which type II theory  either 
a set of vector fields  or a set of antisymmetric tensor naturally
becomes massive. The different cases are summarized in
table~\ref{results}. Of course it is always possible to rotate to a
symplectic basis where all massive modes are either vectors or
tensors.
The most appropriate formulation of the supergravity which occurs
as the low-energy effective theory for the case at hand is the one given in 
ref.~\cite{dWST}. Here all vectors and tensors are kept simultaneously
and the symplectic covariance of the theory becomes manifest.
A reformulation of the results obtained here in terms of the formalism
of \cite{dWST} will be presented elsewhere.

\begin{table}[h]
\begin{center}
\begin{tabular}{| c | c |c|} \hline
   \rule[-0.3cm]{0cm}{0.8cm} 
& IIA& IIB \\ \hline  
\rule[-0.3cm]{0cm}{0.8cm} 
$e_{AI}$ & massive $A_\mu^A$ & massive $A_\mu^I$ \\ \hline
\rule[-0.3cm]{0cm}{0.8cm} 
$\p_A^I$ & massive $A_\mu^A$ & massive $\tilde C_{2}^A$ \\ \hline
\rule[-0.3cm]{0cm}{0.8cm} 
$\m^A_I$ & massive $\tilde C_{2}^I$ & massive $A_\mu^I$ \\ \hline
\rule[-0.3cm]{0cm}{0.8cm} 
$q^{AI}$ & massive $ C_{2  I}$ & massive $ C_{2  A}$ \\ \hline
\end{tabular}
\caption{\small 
\textit{Physical effect of different charges.}}
\label{results}
\end{center}
\end{table}

Finally we come to the issue of mirror symmetry. Comparing
Tables~\ref{N=2multipletsA} and~\ref{N=2multipletsB} results in a
condition purely on the light spectrum. 
First of all the dimensions of the finite subspaces defined in \eqref{deffin}
have to agree on a mirror pair of six-manifolds $(M^6,\tilde{M}^6)$ or in
other words $\beve(M^6)=\bodd(\tilde{M}^6)$ and vice
versa. Furthermore the kinetic terms in the Lagrangian have to
coincide. Here we only computed explicitly the K\"ahler potential of
the two K\"ahler geometries in 
\eqref{Kpmgen}. 
We see that mirror symmetry requires the identification  \cite{FMT}
\begin{equation}\label{mirrorcomp}
\Phi^+(M^6) \leftrightarrow \Phi^-(\tilde{M}^6)\ , \qquad 
\Phi^-(M^6) \leftrightarrow \Phi^+(\tilde {M}^6)\ , 
\end{equation}
or equivalently  the exchange
\begin{equation}
X^A \leftrightarrow Z^I\ , \qquad F_A\leftrightarrow \cF_I\ .
\end{equation}
Comparing also the kinetic terms for the RR scalars is straightforward
and results in the identification 
\begin{equation} 
\qquad
\xi^A \leftrightarrow \xi^I\ , \qquad
\tilde\xi^A \leftrightarrow \tilde\xi^I\ .\qquad
\end{equation}
Finally comparing the Killing prepotentials \eqref{Pdef2},
\eqref{P3def2}
with \eqref{PdefIIB2},
\eqref{P3defIIB2}
requires an identification of the charges
\begin{equation} 
e_{AI}  \leftrightarrow e_{IA}\ , \qquad
q^{AI}\leftrightarrow q^{IA}\ , \qquad
\p^{I}_A \leftrightarrow - \m^A_I\ ,
\end{equation}
and the RR-fluxes
\begin{equation} 
G_{RR}^A \leftrightarrow - G_{RR}^I\ , \qquad 
\tilde G_{RR\, A} \leftrightarrow - \tilde G_{RR\, I}\ .
\end{equation}
Thus we see that within the class of compactifications on
manifolds with $\SU(3)\times\SU(3)$ structure mirror symmetry can be
realized. 

The final task of this paper is to ask to what extend the
compactifications just discussed correspond to \emph{bona fide}
geometrical backgrounds. In particular, can one always find geometries
with truncations satisfying~\eqref{diffQ}, and, if not,
how does this connect to the discussion in
the recent literature. 


\section{Non-geometric backgrounds} 
\label{sec:nongeo}


In our discussion thus far, we have simply assumed that there are
suitable \stt\ manifolds with truncations satisfying the differential
conditions~\eqref{diffcond} in the case of the dual of $H_3$-flux, or,
more generally, conditions~\eqref{diffcond2}. In the following, we
will argue that this is generically not the case. 
Instead, following recent ideas generalizing the notion of a string
background, one must consider ``non-geometrical''
compactifications~\cite{MR}--\cite{IE}. 

The classic examples~\cite{DH,Kachru:2002sk,nongeo} of such backgrounds
are tori, and orbifolds thereof, with NS three-form fluxes and the
corresponding backgrounds related by successive T-duality
transformations. Some of these backgrounds were 
shown to be non-geometric~\cite{Hull}. The corresponding effective
theories were discussed in~\cite{hull,STW}. In refs.~\cite{MR} it was
argued that these  backgrounds correspond to non-commutative (and
non-associative) geometries. The relation between these different view
points has recently been clarified in ref.~\cite{pascalsakura}. 
Note also that essentially two types of non-geometrical
backgrounds have been identified: those which are locally geometrical
but have no sensible global geometrical description; and those which
are not even locally geometrical~\cite{nongeo2,STW2}. Specific
examples of the former type can be realised using the concept of a T-fold,
introduced in ref.~\cite{Hull}. These backgrounds locally look
like manifolds but the transition functions between local
patches are generalised to include T-duality transformations. 

Let us first give a suggestive argument as to why geometrical
compactifications are not sufficient to realize all the charges
in~$\mathcal{Q}$. Suppose for this discussion that the
relations~\eqref{diffQ} are exact and not up to terms which vanish
under the symplectic pairing~\eqref{symplecticbasis}. Given that the
exterior derivative maps $p$-forms to $(p+1)$-forms, we find that,
whatever truncation we choose, the charge 
matrix $\mathcal{Q}$ defined in~\eqref{diffQ} cannot be completely
generic. This suggests that in order to generate all the allowed
elements in $\mathcal{Q}$ one must consider non-geometrical
compactifications. The argument is a follows. Recall that $\Phi^\pm$
are expanded in terms of truncation bases $\Ae$ and $\Ao$ as in~\eqref{Phiexp}. 
From~\eqref{exppurespinors} we see that, whenever $c_\| \neq 0$, the
structure $\Phi^+$ contains a scalar. This implies that at least one of
the forms in the basis $\Ae$ contains a scalar. Let us call this
element $\Ae_1$, 
and take the simple case where the only non-zero elements of ${\cal
  Q}$ are those of the form ${\cal Q}_{\hI}{}^1$ 
(where $\hI=1,...,2\bodd+2$). Thus
$\dd\Ao_{\hI}=\mathcal{Q}_{\hI}{}^1\Ae_1$ and so if
$\mathcal{Q}_{\hI}{}^1\neq0$ then $\dd\Ao_{\hI}$ contains a
scalar. But this is not possible if $\dd$ is an honest exterior
derivative, acting as $\dd: \Lambda^p \to \Lambda^{p+1}$. The same is
true if $c_\|$ in (\ref{exppurespinors}) is zero. In this case, there
may be no scalars in any of the even forms $\Ae$, and for an ``honest''
$\dd$ operator, there should be then no one-forms in $\dd\Ae$. But we
again  see from (\ref{exppurespinors}) that $\Phi^-$ contains a
one-form, and as a consequence so do some of the elements in $\Ao$. 

One way to generate a completely general charge matrix $\mathcal{Q}$
in this picture is to consider a modified operator $\dd$ which is now
a generic map $\dd:\vt^+\to\vt^-$ which satisfies $\dd^2=0$ but 
does not transform the degree of a form properly. In particular it can
map a $p$--form  to a $(p-1)$--form. Of course, $\dd$ does not act 
this way in conventional geometrical compactifications. One is thus
led to conjecture that to obtain a generic $\mathcal{Q}$ we must
consider non-geometrical compactifications. One can still use the
structures~\eqref{diffQ} to derive sensible effective actions,
expanding in bases $\Ae$ and $\Ao$ with a generalised $\dd$ operator,
but there is of course now no interpretation in terms of differential
forms and the exterior derivative. 

As a concrete simplified example of the general ideas discussed above
we consider the case of a reduction on $T^6$ with
$H_3$-flux and the related twisted tori and T-dual compactifications,
following~\cite{DH,Kachru:2002sk,nongeo,Hull,hull}. Collectively we
refer to such compactifications as ``generalised twisted tori''. We
will introduce \stt\ structures on classes of 
these backgrounds and calculate the corresponding charge matrices
$\mathcal{Q}$. More generally, refs.~\cite{STW} (see also
\cite{STW2,Sdual}) looked at $N=1$ orientifolds of such backgrounds,
calculating the corresponding effective superpotentials. In this
subsection we will review the structure of these generalised $T^6$
reductions. In the following subsection we calculate the corresponding
$\mathcal{Q}$ matrices for our putative \stt\ structures and finally
in the last subsection we compare with the superpotential of
ref.~\cite{STW}. 

\subsection{Generalised twisted tori}

A Calabi-Yau manifold in the SYZ limit can be viewed as a three-torus
$T^3$ fibred over some base manifold~\cite{SYZ}. In this limit mirror
symmetry acts as T-duality on the $T^3$ fibre while leaving the base
unchanged. With this prescription one can explicitly construct the
mirror duals of a Calabi-Yau manifold with three-form flux $H$. The
$T^6$ examples we discuss here are the trivial case of such a construction. 

Let us start with a $T^6$ compactification where $e^a$ are a set of
one-forms defining the torus and where we include NS flux
$H=\frac{1}{6}H_{abc}e^a\wedge e^b\wedge e^c$. The action of T-duality
in this background has been considered by many authors. Heuristically,
following the notation of ref.~\cite{STW}, it can be represented
as follows  
\beq\label{fluxchain}
H_{abc} \xleftarrow{\, T_a} \hspace{-8pt} \rightarrow f^a{}_{bc}
\xleftarrow{\, T_b} \hspace{-8pt} \rightarrow Q^{ab}{}_c
\xleftarrow{\, T_c} \hspace{-8pt} \rightarrow  R^{abc}\ .
\eeq
In the SYZ formulation the different terms in~\eqref{fluxchain}
correspond to the situation where $H$ has one, two or three
`legs' on the $T^3$-fibre. An $H$ with one leg on the fibre
corresponds to electric NS-fluxes and has already been considered 
in~\cite{GLMW}. This leads to a geometry described by the parameters
$f^a{}_{bc}$, and no $H$-flux. Geometrically we have a
twisted torus. This is a parallelisable manifold spanned by one-forms
$e^a$, which are now not closed, but satisfy instead 
\beq\label{de}
de^a=f^a{}_{bc} \, e^b \wedge e^c\ ,
\eeq
with $f^a{}_{bc}$ constant. Specifically, suppose only one element of
$H_{abc}$ is non-zero, and has only one leg on the $T^3$
fibration. After three T-dualities, we get a new manifold which is a
non-trivial $T^3$ fibration. The non-trivial part is a $S^1$ fibration
over $T^2$, where the $S^1$ is the T-dual of the fibre direction along
which $H$ was non-zero.  

Now suppose $H$ has two legs along the $T^3$ fibration. One can again
explicitly perform a local T-duality leading to a background with
non-trivial geometry and $H$-flux. However, this cannot be done
globally: there is no good global splitting between metric and
$B$-field. Instead, one can interpret the non-trivial part of the
compactification as a $T^2$ bundle over $S^1$ where there is mondromy
that mixes the $B$-field and metric of the $T^2$: the bundle is being
patched by an element of T-duality. As such it is a T-fold and is
non-geometric. Nonetheless, the reduction can be characterized by a
set of parameters $Q$ which are related to the local metric and
$B$-field. 

Finally, the last step in the chain~\eqref{fluxchain} is purely
conjectural, since the metric does not have the isometry to perform
such T-duality, and therefore the Buscher rules cannot be applied. It
corresponds to an $H$-flux with all three legs on the fibre.  
In this case, \cite{nongeo2} argues that there is not even a good local
description of the geometry, though it does make sense as a conformal
field theory. One way~\cite{STW} to see that space-time points might
not be well defined, is to note that the mirrors of D0-branes probes
would be D3-branes wrapping a $T^3$ fibre with NS flux on the world-volume and
these do not have simple moduli spaces because of the problem of
satisfying the Bianchi identity $\dd F=H_3$. In this sense, the
parameters $R$ have no geometrical interpretation. Note that by an
abuse of nomenclature, we will often refer to all the parameters $H$,
$f$, $Q$ and $R$ as generalised ``fluxes''. 

There are various ways to view what is encoded in these generalised
fluxes. In terms of the corresponding low-energy effective theory they
are related to the gauge algebra of the vector fields, coming from the
symmetries of the backgrounds. One
finds~\cite{KM,DallaFerrara,hull,DH,Hull,nongeo2,STW}    
\beq
\begin{aligned} 
\label{o66algebra}
   [v_a,v_b]=& H_{abc} X^c+ f^c{}_{ab} v_c , \\
   [v_a,X^b]=&-f^b{}_{ac}  X^c+ Q^{bc}{}_{a} v_c ,\\
   [X^a,X^b]=& Q^{ab}{}_c X^c+ R^{abc} v_c ,
\end{aligned}
\eeq
where in the case of a geometrical compactification ($Q=R=0$) the
$v_a$ generators come from the Killing vector symmetries, while $X^a$
are associated with gauge transformations of $B$. Note that the
algebra of diffeomorphisms parametrized by vectors and gauge
transformations parametrized by one-forms is essentially the same as
the Courant bracket\footnote{The Courant bracket between two elements
  $x+\xi$ and $y+\eta$ in $E$ is given by
  $[x+\xi,y+\eta]=[x,y]+\mathcal{L}_x\eta-\mathcal{L}_y\xi
  -\frac12\dd(i_x\eta-i_y\xi)$ where $[x,y]$ is the usual Lie bracket
  of vector fields and $\mathcal{L}_x$ is the Lie derivative.}. From
this perspective, in the geometrical 
case, one can view~\eqref{o66algebra} as the Courant bracket algebra
of Killing vectors and gauge transformations. 
Since, for instance, the gauge transformation of $B$ are Abelian, one
can see that the $Q$ and $R$ fluxes cannot arise in any convention
geometrical way. Note that the Jacobi identities for the algebra then
put constraints on fluxes. 

An alternative picture is that the corresponding generalised geometry
can be written in terms of a basis $V^A$ of $O(6,6)$ vectors, just as for a
twisted torus there is a basis of left-invariant one-forms $e^a$, or
equivalently vectors $\tilde{v}_a$. Just as the structure constants
$f^a{}_{bc} $ appear in the Lie algebra of the $\tilde{v}_a$, so the
generalised fluxes appear in the Courant bracket algebra of the
$V^A$. Note that this is a complementary picture to the one just
given: on a twisted torus the right-invariant vector fields $v_a$
generate the isometries, while the left-invariant vector fields 
$\tilde{v}_a$ are used to define the metric. 

A third picture, useful when relating to \stt\ structures is to ask
how the fluxes enter the exterior algebra of the forms. For a
geometrical background it is natural to consider forms of the type
$\omega=\ee^{-B}\omega_{m_1\dots m_p}e^{m_1}\wedge\dots\wedge
e^{m_p}$ with $\omega_{m_1\dots m_p}$ constant. We include the
twisting by $B$ so that $\omega$ is an element of the generalised
spinor bundle $S(E)$. Acting with $\dd$ on $\omega$ we find
\begin{equation}
   \dd\omega = - H \wedge \omega + f\cdot\omega
\end{equation}
where $(f\cdot\omega)_{m_1\dots m_{p+1}}=
f^a{}_{[m_1m_2|}\omega_{a|m_3\dots m_{p+1}]}$. The natural
non-geometrical extension is then to an operator $\mathcal{D}$ such
that~\cite{STW}
\beq\label{newd}
\cD\omega := - H \wedge\omega + f\cdot\omega + Q \cdot\omega 
   + R \llcorner \omega,
\eeq
where $Q \cdot$ and $R \llcorner$ are defined by
\beq
(Q \cdot \omega)_{m_1 ... m_{p-1}}= Q^{ab}{}_{[m_1}
\omega_{|ab|m_2...m_{p-1}]} \ , \quad 
(R \llcorner \omega)_{m_1...m_{p-3}}= R^{abc}
\omega_{abcm_1...m_{p-3}} \ . 
\eeq
Requiring $\mathcal{D}^2=0$ implies that same conditions on fluxes as
arose from the Jacobi identities for~\eqref{o66algebra}. 
The connection $\mathcal{D}$ appears in the  Bianchi identities
for the RR fluxes, which  in the presence of geometric and
non-geometric fluxes read $\cD F=0$. Note that in our analysis the
equality in~\eqref{newd} will be relaxed to an equivalence up to terms
vanishing under the symplectic pairing~\eqref{symplecticbasis}. 

\subsection{Generalised twisted tori and \stt\ structures}

We will now try and relate the fluxes~\eqref{fluxchain} in the
generalised twisted tori examples to our generic \stt\ reductions
discussed in section~\ref{general}. This will allow us to see how the
charges $\mathcal{Q}$ can be realised in terms of the fluxes and
hence, in this particular example, which terms in $\mathcal{Q}$ come
from conventional compactifications and which from non-geometrical
backgrounds. 

Let us consider first an $\SU(3)$ structure on the generalised twisted
torus manifold. In the geometrical case, the manifold is
parallelisable and there is non-trivial $H$-flux. To define the
$\SU(3)$ structure we introduce three complex one-forms $e^i$
(with conjugates $\bar{e}^{\bar{i}}$). In order to keep the discussion
tractable we will assume that there is $\bbZ_3$ symmetry under
permutation of the three $e^i$. In the simple case where the manifold
is $T^6$ this implies that we are considering the product $T^2 \times
T^2 \times T^2$ and assuming the metric and $H$-field are the same on
each $T^2$.  

In terms of $\SU(3)$ structure this means we fix identical complex
structures and K\"ahler forms on each $T^2$ (or rather in terms of
each $e^i$). There are then two moduli: the complex K\"ahler modulus
$t$ and complex structure $\tau$ of each $T^2$. We thus have, as in
section~\ref{sec:H3} 
\begin{equation}
\label{T6Phi}
   \Phi^+ = \ee^{-\Bfl}\ee^{\ii t \lambda} \ , \qquad 
   \Phi^- = \ee^{-\Bfl}\Omega_\tau^1\wedge\Omega_\tau^2\wedge\Omega_\tau^3 ,
\end{equation}
where $\lambda=2\ii\delta_{i\bar{j}}e^i \bar{e}^{\bar{j}}$ and
$\Omega_\tau^i=\frac12(1+\tau)e^i+\frac12(1-\tau)\bar{e}^{\bar{i}}$ define the
complex structure on each $T^2$, while $\dd\Bfl=H$. We are expanding
in a basis of even forms 
\begin{equation} 
\label{Aev}
   \Ae = (\omega_0,\omega_1,\tilde\omega^{0},\tilde \omega^{1})
      = \ee^{-\Bfl}\left(
         1, \tfrac16 \lambda^2, \tfrac16 \lambda^3, \lambda\right) , 
\end{equation}
and of odd forms
\begin{equation}
\label{Aodd}
   \Ao = (\alpha_0, \alpha_1, \beta^{0}, \beta^{1})
      = \ee^{-\Bfl}\left(\re\Omega_3, \re\chi_3, -\im\Omega_3,
         -3\im\chi_3 \right)  
\end{equation}
where
\beq
\Omega_3 = \tfrac{2}{3} \epsilon_{ijk} e^i e^j e^k \ , \qquad 
\chi_3  = \tfrac43 (\bar{e}^{\bar 1}e^2e^3 + \cycle)   
   =: \tfrac{2}{3}\, \rho^i{}_{jk} \, 
      \delta_{i \bar l} \, \bar{e}^{\bar l} e^je^k \  .
\eeq
The components of $\rho$ satisfy  $\rho^1{}_{23}=-\rho^1{}_{32}=
\rho^2{}_{31}=-\rho^2{}_{13}=\rho^3{}_{12}=-\rho^3{}_{21}=1$, with the
others being zero. The forms satisfy
additionally~(\ref{compbasis}). 

The fluxes~\eqref{fluxchain} of the non-trivial geometry are
encoded in the $H$-flux and the twisted
geometry~\eqref{de}. Specifically, respecting the $\bbZ_3$ symmetry we  
have  
\begin{equation} 
\label{Hsu3}
   H_3 = \dd \Bfl =
      H^0 \re \Omega_3 + H^1 \re \chi_3 - H_0 (-\im\Omega_3) 
         - H_1 (-3\im\chi_3) \ ,
\end{equation}
while decomposing~\eqref{de} in terms of holomorphic and
antiholomorphic indices, and imposing the $\bbZ_3$ symmetry, gives 
\begin{equation} 
\label{strcsu3}
   \dd e^i = \tfrac{1}{6}A\rho^i{}_{jk} \, e^j e^k 
      + \tfrac{1}{6}B\rho^{ij}{}_{k} \, 
         \delta_{j \bar l}\, \bar{e}^{\bar l} e^k  
      + \tfrac{1}{6}C\epsilon^{ijk} \, 
         \delta_{j \bar l}  \, \delta_{k \bar m } \,
         \bar{e}^{\bar l} \bar{e}^{\bar m} .  
\end{equation}

Using~(\ref{Hsu3}) and~(\ref{strcsu3}) to compute the exterior
derivatives of the elements of $\Ae$, and expressing them as linear
combinations of the forms in $\Ao$ we obtain an expression for the charge
matrix ${\cal Q}$ in terms of the structure constants $A$, $B$ and $C$
and the $H$-fluxes $H^I$ and $H_I$. We get  
\begin{equation} \label{Hf}
\begin{gathered}
   H_I = e_{I0} \ , \qquad H^I=\p^I{}_0 \ ,\\
   6A + \bar B = 3 p_1{}^1+\ii q^{11} \ , \\
   C = \tfrac{1}{6}p_0{}^1+\tfrac{1}{6}\ii q^{01} \ ,
\end{gathered}
\end{equation}
and $q^{I0}=p_I{}^0=e_{I1}=m^I{}_1=0$.\footnote{Note that for our
  choice of $SU(3)$ structure not all fluxes of \eqref{strcsu3} appear
  but only the combination $\bar B+ 6 A$.}
The charge matrix is therefore
\beq \label{Qsu3geo}
{\cal Q}= \left(\begin{array}{cccc}
0 &  \RE C & H_1 & 0\\
0 & \frac1{18} \RE D  & H_2 & 0\\
0 &  \IM C & H^1 & 0\\
0 &  \IM D & H^2 & 0
\end{array} \right) \ , 
\eeq
where $D=6A+\bar{B}$. 
This implies that only half of the charges are turned on via H-flux
and geometric fluxes.
We therefore expect the other half of the charges ${\cal Q}_{\hI 1}$,
${\cal Q}_{\hI 4}$ ($\hI=1,...,4$) to correspond to non-geometric
fluxes. There are as many $Q^{ab}{}_c$  fluxes respecting the
${\mathbb Z}_3$ symmetry as there are $f^a{}_{bc}$, and the same is
true for $R^{abc}$ and $H_{abc}$. It is reasonable to expect  that
turning them on would complete the entries of the charge matrix ${\cal
  Q}$. Let us show that this is indeed the case. 

Let us use the operator $\cD$ in~(\ref{newd}) to define the fluxes $Q$
and $R$. Replacing $\dd$ in~\eqref{diffcond2} with $\mathcal{D}$ we
find that the full charge matrix is then given by
\beq
\label{QR}
{\cal Q}= \left(\begin{array}{cccc}
R_1 &  \RE C & H_1 & \frac23 \IM \tilde C \\
R_2 & \frac1{18} \RE D  & H_2 & \frac{2}{9}  \IM \tilde D \\
R^1 &  \IM C & H^1 & \frac23 \RE \tilde C \\
R^2 &  \IM D & H^2 &  \frac16 \RE  \tilde D
\end{array} \right) \ , 
\eeq
where $\tilde{D}=\tilde{A}+\bar{\tilde{B}}$ and $\tilde{A}$,
$\tilde{B}$ and $\tilde{C}$ are defined in direct analogy with $A$,
$B$ and $C$, while
 $R_I$, $R^I$ are the components of $R$-flux
defined in analogy with~\eqref{Hsu3}. 
We see, as promised,  that the missing half of the ${\cal Q}$'s are
indeed given by the non-geometric fluxes $Q$ and $R$. We conclude that
the charge matrix ${\cal Q}$ represents geometric as well as
non-geometric fluxes, and all of the elements of ${\cal Q}$ can in
principle be generated by an appropriate $H$, $f$, $Q$ or
$R$-flux. Note that the flux parameters  are not all independent but
have to satisfy the constraint~\eqref{diffconstraints}. The same
constraint also arises from requiring $\cD^2=0$. In this particular
case, using the general expression~\eqref{dimQ}, we have ten
independent charges. 

We can also generalize this calculation to the case of an \stt\
structure. From the discussions in the previous sections, we expect
this setup to accommodate more of the $\mathcal{Q}$ charges in a
purely geometric background. We will see that this is indeed the case.  

Specifically we assume that there is an $\SU(2)$ structure
on the generalised $T^6$ again with ${\mathbb Z}_3$ symmetry. Using
the same forms $e^i$, let us choose $e^3$ to be the holomorphic 
vector of the $\SU(2)$ structure. In the language of
Eq.~(\ref{exppurespinors}), we are taking $c_\|=0$, $c_\bot=1$ and
$v+\ii v'=e^3$. The $\SU(2)$ structure is then equivalent to two
$\SU(3)$ structures, defined by the holomorphic vectors $(e^1,e^2,e^3)$
and $(\hat e^1,\hat e^2,\hat e^3)=(\bar{e}^{\bar 1},\bar{e}^{\bar 2},e^3)$. The
${\mathbb Z}_3$ acts by a simultaneous permutation of $(e^1,e^2,e^3)$
and $(\hat e^1, \hat e^2, \hat e^3)$. We can again find suitable
bases $\Ae$ and $\Ao$ preserving the $\bbZ_3$ symmetry and
(\ref{symplecticbasis}) and (\ref{compbasis}). The bases with the
minimum number of elements are given by  
\begin{equation*}
\Ae= \ee^{-\Bfl}\left( \begin{array}{c}
      2\, \RE(\omega_2 + \xi_2) \\
      8\, \IM \omega_2 -  4\, \ii \RE(\omega_2 + \xi_2) e^3 \bar e^3 \\
      -4 \ii \RE(\omega_2-\xi_2) e^3  e^{\bar 3} \\
      -\frac23  \RE(\omega_2-\xi_2) + \frac23
      \IM\omega_2 e^3 e^{\bar 3}
   \end{array} \right) \ , \quad
\Ao= \ee^{-\Bfl} \left( \begin{array}{c}
      2 \RE e^3 \\ -2 \IM e^3 +  \RE e^3 j\\ - \IM e^3 j^2 \\ -\frac13
      \RE e^3 j^2  +\frac43 \IM e^3 j 
   \end{array}  \right)\ ,
\end{equation*}
where wedge products are understood and where $\omega_2=e^1\wedge
e^2$, $\chi_2= \bar{e}^{\bar 1}\wedge e^2$, and $j=2\ii (e^1\wedge
\bar{e}^{\bar 1}+ e^2\wedge  \bar{e}^{\bar 2})$. Note that,  
with $\Bfl=0$, there are neither scalars, nor six-forms in the basis of even
forms. In addition, unlike in the $\SU(3)$ case with $\Bfl=0$, it is not
possible to find a basis of forms of pure degree.
 
The ``metric fluxes'' are introduced via the exterior derivatives of
the one-forms, given by~(\ref{strcsu3}). In the symmetric setup, the
structure constants are again proportional to $\epsilon^{ijk}$ and
$\rho^i\,_{jk}$. As before the $H$-flux, comes from the twisting of
the basis forms by $e^{-\Bfl}$. Since there are no scalars in the
basis of even forms, we should not expand $H_3$ in the basis of odd
forms, but rather simply calculate the parameters $H_{\hat I \hat
  A}=\int \mukai{H_3 \wedge \Ao_{\hat I}}{\Ae_{\hat A}}$. 

The structure constants and $H$-flux generate the following charge matrix 
\beq \label{qsu2}
{\cal Q}= \left( \begin{array}{cccc} 
\frac1{12} \RE E^+ & \frac{1}{72} \IM F + h^+_i & 3 h^-_i &
-\frac16\RE E^-+4 h^0_i \\ 
-\frac{1}{12} \IM E^+ & \frac{1}{24} \RE F+h^+_r & \frac{1}{12} \IM
E^-+3 h^-_r  
&  \frac16 \IM (2E^-+F) +4 h^0_r \\
0 & 0 & 0 & 0\\
0 & \frac{1}{108} \IM (3 E^+-E^-) & \frac19  \IM F & -\frac{2}{9}
\RE(E^+ + F) 
\end{array}
\right) \ , \nn 
\eeq
where we have defined
\beq
E^\pm=A+C\pm 2B \ , \quad  F=-A+C .
\eeq
The parameters $A$, $B$, and $C$ are defined in (\ref{strcsu3}), and
$h^{\pm,0}_{r,i}$ are the different $H$-flux charges that can be  
turned on. If we expanded $H_3$ in 20 independent three-forms, only six
combinations 
of them would contribute to the charges. Explicitly, 
\begin{equation}
\begin{aligned}
   H_3 &= \tfrac{1}{6} h^\pm_r \RE(\omega_2 \pm \xi_2) \RE e^3 + 
      \tfrac{1}{6} h^\pm_i \RE(\omega_2 \pm \xi_2) \IM e^3 +
      h^0_r \IM(\omega_2) \RE e^3 \\
         &\qquad \qquad + h^0_i \IM(\omega_2) \IM e^3 + \dots
\end{aligned}
\end{equation}
where the $+\dots$ are pieces that do not contribute to the charge
matrix. We see that in the $\SU(2)$ case, 11 out of the 16 charges can
be turned on via geometric fluxes, as oposed to 8/16 for the $\SU(3)$
case. The remaining 5 charges can be turned on by $Q$- and $R$-fluxes. Note
once more that there are (six) conditions on the charges coming from
constraint~(\ref{diffconstraints}). For the charge matrix
(\ref{qsu2}), two of these are automatic, while one needs to impose
the other four.  

We conclude that in order to generate non-zero entries for the full
charge matrix we need geometric as well as non-geometric fluxes both
in the $\SU(3)$ and in the $\SU(2)$ case. However, in the latter the
number of charges that can be turned on via geometric fluxes is
generically larger than in the former. 

  \subsection{Superpotentials}

We can further support the claim that a generic ${\cal Q}$ contains
geometric and non-geometric fluxes by computing  the superpotentials
(\ref{Pdef2}) and (\ref{PdefIIB2}) for a given $\mathcal{Q}$, and
comparing to that of ref.~\cite{STW}. Starting from IIA and IIB
compactifications on the $\mathbb{Z}_3$ symmetric $T^2 \times T^2
\times T^2$ torus with an $\SU(3)$-structure, flux and O6 and O3
planes respectively, the authors of~\cite{STW} used T-duality
arguments to propose a generic form for the superpotential valid also
for dual non-geometrical compactifications.  The superpotentials are
functions of the dilaton $S$, two further $N=1$ moduli $X$ and $Y$
and the fluxes $H$, $f$, $Q$, and $R$. They have the generic form
\beq \label{Wnongeo}
{\cal W}=P_1(X) + S P_2(X) + Y P_3(X) \ ,
\eeq
where $P_{1,2,3}(X)$ are cubic polynomials
with the coefficients being the (geometric and non-geometric) NS
and RR fluxes. $P_1$ depends on RR fluxes only, while the
NS fluxes generate $P_2$ and $P_3$.  Each type of flux contributes
to a term with a given dependence on the moduli. For example, the term
proportional to $S X^2$ is proportional to $Q$-flux in type IIA, while
it corresponds to $H$-flux in type IIB. 

Let us compare (\ref{Wnongeo})  with the superpotential obtained from
the type IIA and type IIB superpotentials given in~(\ref{WA})
and~(\ref{WB}), for an O6 and an O3 orientifold projection
respectively. The $N=1$ supersymmetry preserved by these projections
correspond to  $\alpha=\pi/4$, $\beta=\pi/2$, giving
\begin{align}
\label{WO6}
   && {\cal W}_{\rm{IIA/O6}} &=  \int \mukai{\Phi^+}{\dd \Pi^-} \ , &
   \Pi^- &:= A_0^- + \ii \RE(C \Phi^-) \ , \\
\label{WO3}
   && {\cal W}_{\rm{IIB/O3}} &= - \int \mukai{\Phi^-}{\dd \Pi^+} \ , &
   \Pi^+ &:= A_0^+ + \ii \RE(\ee^{-\phi} \Phi^+) \ ,
\end{align}
where $A_0^\pm$ are the RR potentials defined in \eqref{CAexp},
\eqref{CBexp} with field strength $G^\pm$ defined in \eqref{GAdef}.
In ref.~\cite{iman} it was shown that $\Pi^\pm$ are the correct $N=1$
K\"ahler coordinates for the orientifolds.  $C$ is a `compensator' field proportional to $\ee^{-\phi}$ (for the
precise definition see \cite{iman}).

Recall that for the symmetric $(T^2)^3$ setup, the $\Phi^\pm$
corresponding to a single $\SU(3)$ are given by~\eqref{T6Phi} with
moduli $t$ and $\tau$. 
After the O6 orientifold projection $t$ remains an $N=1$ modulus
(which is commonly called $T$) while $\tau$ is constrained to be real
and it combines with a RR scalar $\xi_1$ to form the $N=1$ modulus $U
=\xi_1 + iC\tau^2$ which enters $\Pi^-$. The second variable is
$S=\xi_0 + i C$. In type IIB, the O3 projection requires $t$ to be
real, and the $N=1$ moduli are given by $U=\tau$, $T=\xi_1 + \ii
\ee^{-\phi} t^2$ and $S=\xi_0 + \ii \ee^{-\phi}$ (see~\cite{iman} for
further details). 

Substituting these expressions and using the bases~\eqref{Aev}
and~\eqref{Aodd} and the general expressions~(\ref{diffQ})
and~(\ref{chargem}) we find
\beq
\begin{aligned} \label{Wfinal}
&{\cal W}_{\rm{IIA/O6}} = U\Big[
   \ii (3 e_{00}-e_{10}) - T (3 p_0{}^1 - p_1{}^1)
   -3\ii T^2 (3 e_{01} - e_{11}) - T^3 (3 p_0{}^0 - p_1{}^0) \Big]
\\ & \qquad {}+ S\Big[
   \ii (e_{00}+e_{10}) - T (p_0{}^1 + p_1{}^1)
   -3\ii T^2  (e_{01} + e_{11}) - T^3 (p_0{}^0 + p_1{}^0) \Big] \ .
\end{aligned}
\eeq
for type IIA, and
\beq
\begin{aligned} \label{WfinalB}
&{\cal W}_{\rm{IIB/O3}} = T\Big[
   3 \ii (e_{01}+e_{11}) + U (3 m^0{}_1 + m^1{}_1)
   -3\ii U^2 (3 e_{01} - e_{11}) - U^3 ( m^0{}_1 - m^1{}_1) \Big] 
\\ & \qquad {}+ S\Big[
   - \ii (e_{00}+e_{10}) + U (m^1{}_0 + 3 m^0{}_0)
   -\ii U^2  (3 e_{00} - e_{10}) + U^3 (m^0{}_{0}+m^1{}_{0}) \Big] \ .
\end{aligned}
\eeq
for type IIB. 

These superpotentials are symmetric under the mirror map (\ref{mirrorcomp}).
Furthermore, they contain all the terms in~(\ref{Wnongeo}) 
depending on NS fluxes, namely $P_2$ and $P_3$, if we identify $X=T$
and $Y=U$ for type IIA, and $X=U$ and $Y=T$ for type IIB. The first
lines of (\ref{Wfinal}) and (\ref{WfinalB}) correspond to the
terms in $P_2$, while the second line to those in $P_3$. In the IIA
expression of ref.~\cite{STW}, the terms with no power of $T$
(appearing first on the first and second lines of (\ref{Wfinal}),
proportional to $e_{I0}$) come from $H$-flux. The terms linear in $T$
come from $f$-fluxes and the ones quadratic in $T$ from $Q$-fluxes,
while the cubic ones involve the conjectured $R$-fluxes. This is in
perfect agreement with (\ref{Hf}) and (\ref{QR}), where we identified
$e_{I0}$ charges as $H$-flux, $p_I{}^1$ as $f$-flux, $e_{I1}$ as $Q$-flux
and  $p_I{}^0$ as $R$-flux. Note that the fluxes $m$ and $q$ drop from
the IIA/O6 superpotential (or more precisely, they are projected out by
the orientifold projection). In type IIB with an O3 projection, all
the terms containing the modulus $S$ correpond to $H$-fluxes, while
the ones with a $T$ modulus are generated by $Q$-fluxes. ($f$ and $R$
fluxes are not allowed by an O3 projection.) This is again consistent
with (\ref{Hf}), (\ref{QR}) where $m^I{}_0$ has been identified with
$H^I$, while $m^I{}_1$ with $Q$-flux. 

From these examples, we conclude that the general matrix ${\cal Q}$
contains all possible NS fluxes. Note that the mapping between the
charges $(e,m,p,q)$ and the fluxes $(H,f,Q,R)$  depends on the choice
of basis~\eqref{Aev} and~\eqref{Aodd}. 
However, the fact  that some of these fluxes cannot be obtained from
an honest exterior derivative (or from purely geometric fluxes) is a
basis independent statement.   

\bigskip
\bigskip

The form of the generalised derivative~\eqref{newd} suggests that both
$Q$ and $R$ fluxes are associated with deformations of the usual
exterior algebra. However, we also know that backgrounds with
non-trivial $Q$-fluxes are still locally geometrical. The non-geometry
only appears globally. Thus one might still expect the exterior
algebra to be undeformed working on a patch. A possible resolution is
that~\eqref{newd} is too strong for two reasons. First it gives the
action of $\mathcal{D}$ on forms of pure degree, whereas we have
already seen generically we are interested in basis forms of mixed
degree. Secondly, for our~\stt\ structure we also only 
require an equivalence ``$\sim$'' up to terms which vanish under the
symplectic pairing~\eqref{symplecticbasis}. It would be interesting to
clarify if the exterior derivative actually needs to be modified to
define $Q$ given these two subtleties. For now, let us simply connect
the analysis here to the discussion in~\cite{pascalsakura}, which will 
provide some evidence that such a resolution is possible.  

In section~\ref{sec:H3} we observed that the effect of the $H$-flux
was to twist the geometrical basis of forms so that, for instance,
$\omega=\ee^{-\Bfl}\omo$, which were forms of mixed degree. It is
natural to ask if, for instance, the $Q$-charge can also be realised
as a twisting of the geometrical basis, again giving forms of mixed
degree. This can indeed be done, but the price to pay is higher than
for $H$. Under two T-dualities along the $B$-field directions, the
$B$-transform is mapped to a $\beta$-transform~\cite{pascalsakura} 
(see also~\cite{MPZ}), where $\beta^{ab}$ is a bivector along the
T-dualized directions. Defining a new basis
$\omega=\ee^{\beta\llcorner}\omo$ one would then expect that the
corresponding exterior algebra encodes the $Q$-charges, without
modifying the $\dd$ operator. This is fine locally but globally the
geometrical picture breaks down. Non-trivial $H$-flux corresponds to
patching the bundle $E$ with non-trivial transformations
$B_\alpha=B_\beta+\dd A_{\alpha\beta}$ on the intersection
$U_\alpha\cap U_\beta$. The pure spinors $\Phi^\pm$ are global
sections of the twisted spin bundle $S(E)$. 
In the case of a torus fibration with $H$-flux there are
$B$-transformation monodromies on the $T^3$ fibre as one transverses a
loop in the base. However, since $\Phi^\pm$ are global sections they
are invariant under these monodromies. For the dual $T^3$-fibred
background, the patching is by $\beta$-transformations, that is
T-dualities on the $T^3$ fibres. Such a background is thus not
globally geometrical. There are T-duality-valued monodromies, which
have, for instance, the effect of changing the dimension of a
brane~\cite{Hull,pascalsakura} and the type $k$ of a pure
spinor\footnote{A pure spinor can aways be written as $ 
  \ee^A \theta^1\wedge\dots\wedge\theta^k $, where $A$ is a complex
  two-form and $\theta^i$ are complex one-forms. The integer $k$ is
  the "type" of the pure spinor. }.  
However, the new background still leads to a supersymmetric effective
action, which means there is still a notion of a global \stt\
structure. In other words there is a unique pair of pure spinors
$\Phi^\pm$ on each local geometrical patch. In going between patches
these are related by T-duality transformations, in such a way that
they are invariant under the monodromies.  Expanding in terms of basis
forms $\Ae$ and $\Ao$, this implies that each element of the basis should
similarly be globally defined in this generalised ``bundle'' patched
by T-duality. The usual exterior derivative acting on the basis
elements on each local geometrical patch should encode the $Q$-fluxes,
and the local expressions for the superpotential and so on will still
hold. This is one way of suggesting why the geometrical \stt\
expressions give the correct low-energy effective theory in the case
of non-geometrical compactifications with $Q$-flux. 
 
In summary, we have shown that a generic matrix ${\cal Q}$ contains
geometric as well as non-geometric NS fluxes, by calculating
$\mathcal{Q}$ in terms of the fluxes $H$, $f$, $Q$ and $R$ in the
context of generalised twisted-tori. We further show that, in the
orientifold case, this then reproduces the superpotentials given
in~\cite{STW}. Remarkably we note that treating the exterior
derivative operator in~\eqref{diffcond2} as a generalised linear
operator on the bases forms $\Ae$ and $\Ao$ reproduces the conjectured
non-geometrical superpotentials even when the background is not even
locally geometrical.


\section{Conclusions}
\label{conclusion}


In this paper we completed our study of type II compactifications on
manifolds
with \stt\ structure by further generalizing the formalism developed in
ref.~\cite{GLW}. We first decomposed the ten-dimensional
fields under \stt\  projecting out
all representations $(\rep{3}, \rep 1)$, $(\rep{1}, \rep 3)$ and their
complex conjugates. This corresponds to a reorganization of the
ten-dimensional fields
in terms of `$N=2$ multiplets' without performing a Kaluza-Klein
reduction. In this ten-dimensional framework
we computed the equivalent of the
gravitino mass matrix $S_{AB}$  and the $N=1$ superpotential $W$ for
type IIA and type IIB.
These have the same functional expression in terms of the two pure
spinors
$\Phi^\pm$ and RR field strengths $G^\pm$ as their $\SU(3)$ structure
counterparts found in \cite{GLW}, and are in particular mirror
symmetric under a chirality exchange of the pure spinors and RR fluxes.

We discussed the conditions for a consistent reduction
where the infinite tower of Kaluza--Klein states is truncated to a
set of light modes of the compactification. Such conditions arise from
demanding that the local special K\"ahler geometry of the untruncated
theory descends to the moduli space of truncated modes. (Note the
question of when such truncations exist remains an open problem, see
also~\cite{mkp}.) 
Upon meeting these conditions, the resulting theory
is a four-dimensional $N=2$ supergravity,
with generically massive antisymmetric tensors.
For a specific choice of truncation, we precisely reproduced the
type IIA dual of type IIB supergravity on Calabi-Yau threefolds with
magnetic NS three-form fluxes. This theory was missing in
\cite{GLMW,GLW} but can be found when the compactification manifold
has \stt\ instead of $\SU(3)$ structure. The crucial new ingredient is
the existence of all odd forms including one- and five-forms which are
absent in $\SU(3)$ structure compactifications. 
This allows one to generalise previous Ans\"atze for the exterior derivatives
of the basis forms, involving a doubly symplectic
charge matrix ${\cal Q}$, which encodes
the full set of NS fluxes (three-form flux $H_3$ and torsion).

For general \stt\ structure compactifications the low-energy effective
type IIA and type IIB theories are perfectly mirror symmetric under
exchange of the ``moduli'' $X^A$ and $Z^I$ parameterising the
bundles of even and odd pure spinors (some of these are massive
and therefore not moduli in the strict sense), an exchange of the RR  
fluxes $G_{\rm{RR}\, A}$
and $G_{\rm{RR}\, I}$, and a symplectic transposition
of the charge matrix ${\cal Q}$. The latter maps in particular the
``magnetic'' fluxes
$\p^I{}_A$ to the new set of fluxes $\m^A{}_I$.
The question of the existence of  manifolds of \stt\ structure was not
adressed in this paper. However, the restoration of mirror symmetry
seems to be a strong argument in its favor. 

In spite of the fact that \stt\ structures (or the existence
of one- and five-forms in the basis of odd forms) allow one 
to turn on more components of ${\cal Q}$ than those allowed by pure
$\SU(3)$ structures, we showed that entirely geometric fluxes ($H_3$
plus torsion) do not suffice to generate all components of
${\cal Q}$. 
The extra components were shown to be associated to non-geometric
fluxes, which arise in certain standard cases by performing successive
T-dualities on backgrounds with purely geometric fluxes. A general
charge matrix corresponds to a generic map 
from the truncated space of even forms to the space of odd forms.
In the analysis of~\cite{STW} it corresponds to a generalised
nilpotent operator ${\cal D}=-H\wedge + f \cdot + Q \cdot + R
\llcorner$ acting on the basis of forms. The nilpotency condition
translates into quadratic constraints on ${\cal Q}$ that leave
$(2\beve+3) (\bodd +1)$ (for  $\beve>\bodd$) independent components in
the charge matrix. 

The non-geometrical fluxes $Q$ are associated with a background
which is locally geometrical but globally is patched using T-duality
transformations. As such it can be interpreted as a ``T-fold''
following~\cite{Hull}.  The non-geometrical fluxes $R$ correspond to
backgrounds which are not even locally geometrical. These have been
discussed in~\cite{nongeo2}. In the
former case, supersymmetry implies that one can still identify a local
\stt\ structure. In fact, given that T-duality transformations by
which the background is patched should not break supersymmetry, we
would expect the \stt\ is globally defined, in 
the sense that there are no monodromies. This will not however be true
of the metric and $B$-field, since there is no longer a global
``polarization'' (in the language of~\cite{Hull}). For instance, there
are generically monodromies under which D0-branes become D2-branes and
so on. Remarkably, we find that while derived using the assumption
that we had a geometrical background, our expressions such as that of
the superpotential seem to correctly reproduce the gaugings or masses 
coming from such non-geometric fluxes. The only modification is to
allow a generalised exterior derivative operator or, in the truncated
version, a general charge matrix $\mathcal{Q}$. While in the case of
$Q$ fluxes this might be assumed to be related to the local
geometrical structure, the expressions also appear to hold for
$R$-fluxes where the background is not even locally geometrical.


\subsection*{Acknowledgments}

This work is supported by DFG -- The German Science Foundation, the
European RTN Programs HPRN-CT-2000-00148, HPRN-CT-2000-00122,
HPRN-CT-2000-00131, MRTN-CT-2004-005104, MRTN-CT-2004-503369 and the
DAAD -- the German Academic Exchange Service.  D.W.\ is supported
by a Royal Society University Research Fellowship. M.G. is partially
supported by ANR grant BLAN06-3-137168.

We have greatly
benefited from conversations and correspondence with Gianguido
Dall'Agata, Jerome Gauntlett, Pascal Grange, Chris Hull, Amir
Kashani-Poor, Ruben Minasian, Thomas Grimm,
Sakura Sch\"afer-Nameki, Wati Taylor,
Alessandro Tomasiello, Silvia Vaula and Brian Wecht.

J.L.\ thanks David Gross and the organizers of the KITP workshop
``Geometrical Structures in String Theory'' and M.G.\  thanks
the Institute for Mathematical Sciences at Imperial College
for hospitality and financial support during initial and final stages
of this work.


\appendix


\section{Spinor conventions}
\label{app:spinors}


For convenience, in this appendix we will summarize our conventions
for $O(6,6)$ spinors and identify the various relations to
conventional $\Spin(6)$ representations. We start by defining our
conventions for $\Spin(6)$ spinors. 

\subsection{$\Spin(6)$ spinors}

The Clifford algebra $\Cliff(6,0;\bbR)$ is generated by the gamma
matrices $\gamma_m$ satisfying 
\begin{equation}
   \{\gamma_m,\gamma_n\}=2g_{mn} . 
\end{equation}
where $g$ is a positive definite six-dimensional metric. Let
$\epsilon_g$ be an orientation compatible with $g$ (and thus fixed up
to a sign). We can define the standard intertwiners
\begin{equation}
   \gamma^\dag_m=A\gamma_m A^{-1} , \qquad
   -\gamma_m^T=C^{-1}\gamma_mC , \qquad
   -\gamma^*_m=D^{-1}\gamma_m D ,
\end{equation}
and the chirality operator
$\gamma_{(6)}=\frac{1}{6!}\epsilon_g^{m_1\dots m_6}\gamma_{m_1\dots
  m_6}$. Note one can always choose a representation where $A=C=D=1$
and the $\gamma^m$ are imaginary and anti-symmetric. For a spinor
$\theta$ it is useful to define 
\begin{equation}
   \bar{\theta}=\theta^\dag A , \qquad
   \theta^t = \theta^TC^{-1} , \qquad
   \theta^c = D\theta^* .
\end{equation}
We also define chiral spinors by
$\gamma_{(6)}\theta_\pm=\mp\ii\theta_\pm$ with
$\theta^c_\pm=\theta_\mp$. 

\subsection{$\Spin(6,6)$ spinors}

Let $\Pi,\Sigma,\dots$ denote $O(6,6)$ vector indices on the
generalised bundle $E$. (For simplicity here we will assume $E=F\oplus
F^*$.) The Clifford algebra $\Cliff(6,6;\bbR)$ is generated by the
gamma matrices $\Gamma^\Sigma$ satisfying 
\begin{equation}
\label{Cliff12}
   \left\{ \Gamma_\Pi, \Gamma_\Sigma \right\} 
      = 2\mathcal{G}_{\Pi\Sigma} ,
\end{equation}
where $\mathcal{G}$ is the $O(6,6)$ invariant 
metric~\eqref{sGdef}. The $O(6,6)$ spinors $\chis\in S$ can be chosen
to be Majorana--Weyl and we write $\chis^\pm\in S^\pm$ for the two 
chiralities. As usual one can define the intertwiner   
$-\Gamma^T_\Sigma=\mathcal{C}^{-1}\Gamma_\Sigma\mathcal{C}$. Using
$\mathcal{C}$ one can define a spinor bilinear (which defines the
Mukai pairing) by   
\begin{equation}
\label{bilinear}
   \psis^t\cdot\chis := \psis^T \mathcal{C}^{-1}\chis . 
\end{equation}
Since $\mathcal{C}^T=-\mathcal{C}$ this is actually defines a
symplectic structure. The Majorana condition uses the intertwiner
$\Gamma_\Sigma^*=
\tilde{\mathcal{D}}^{-1}\Gamma_\Sigma\tilde{\mathcal{D}}$, and reads
$\chis^{\tilde{c}}:=\tilde{\mathcal{D}}\chis^*=\chis$. 

There are a number of different sub-groups of $O(6,6)$ under which we
can decompose the spinor representation. First, the decomposition
$E=F\oplus F^*$ defines a $\GL(6,\bbR)\subset O(6,6)$ group. A vector
$V\in E$ can then be decomposed into an ordinary 
vector and one-form $V=x+\xi$. Furthermore, under this map $S$ is
isomorphic to the bundle of forms $S\simeq\Lambda^*F^*$ (or for chiral
spinors $S^+\simeq\Leven F^*$ and $S^-\simeq\Lodd F^*$) 
\begin{equation}
\label{isom}
   \chis \sim \chi = \chi_0 + \dots + \chi_6 , 
\end{equation}
where $\chi_p\in\Lambda^pF^*$ and the isomorphism depends on a choice
of volume form $\epsilon$ (though is independent of the sign of
$\epsilon$). In this basis, the metric $\mathcal{G}$ has the
form~\eqref{sGdef} and we can decompose the gamma matrices as  
\begin{equation}
   V^\Sigma \Gamma_\Sigma = x^m \Gv_m + \xi_m \Gf^m
\end{equation}
so that~\eqref{Cliff12} becomes 
\begin{equation}
   \{ \Gv_m,\Gv_n \} = \{ \Gf^m, \Gf^n \} = 0 , \qquad
   \{ \Gv_m,\Gf^n \} = 2\delta_m^n .
\end{equation}
Under the isomorphism~\eqref{isom}, the Clifford action on $\chi$ is
given by
\begin{equation}
\label{Caction}
   (V^\Sigma\Gamma_\Sigma)\chis \sim i_x\chi + \xi\wedge\chi .
\end{equation}
The spinor bilinear decomposes into the Mukai paring on the
constituent forms
\begin{equation}
   (\psis^t\cdot\chis)\,\epsilon
     = \mukai{\psi}{\chi} 
     = \sum_p (-)^{[(p+1)/2]} \psi_p \wedge \chi_{6-p} . 
\end{equation}

The next subgroup one is interested in is the $O(6)\times O(6)\subset
O(6,6)$ structure on $E$ defined by a choice of metric $g$ and
$B$-field. Specifically in terms of the gamma matrices one can use $g$
and $B$ to change basis
\begin{equation}
\label{G+-}
   \Gamma^\pm_m 
      = \frac{1}{\sqrt{2}}\left(\Gv_m + (B_{mn}\pm g_{mn})\Gf^n\right)
\end{equation}
so the Clifford algebra becomes
\begin{equation}
   \{ \Gamma^+_m,\Gamma^-_n\} = 0 , \qquad
   \{ \Gamma^+_m,\Gamma^+_n \} = 2g_{mn} , \qquad
   \{ \Gamma^-_m,\Gamma^-_n \} = - 2g_{mn} . 
\end{equation}
In this basis $\mathcal{G}$ is block diagonal. Clearly $\Gamma^\pm$
generate two different $\Spin(6)$ subgroups. We can correspondingly 
decompose the Clifford algebra
$\Cliff(6,6;\bbR)\simeq\Cliff(6,0;\bbR)\times\Cliff(6,0;\bbR)$. The
spinor bundle is then a product $S=S_1\otimes S_2$ with 
$\chis=\theta_1\otimes\theta_2$ and gamma matrices 
\begin{equation}
   \Gamma^+_m = \gamma_m \otimes \id , \qquad
   \Gamma^-_m = \gamma_{(6)} \otimes \gamma_m ,  
\end{equation}
where $\gamma_m$ are defined above. The intertwiners
$\mathcal{C}$ and $\tilde{\mathcal{D}}$ are given by
\begin{equation}
\label{interdecomp}
   \mathcal{C}=C\otimes C\gamma_{(6)} , \qquad
   \tilde{\mathcal{D}} = D\gamma_{(6)}\otimes D\gamma_{(6)} . 
\end{equation}
The $O(6,6)$ chirality operator is given by 
\begin{equation}
\label{chdef}
   \Gamma_{(12)} = - \gamma_{(6)}\otimes \gamma_{(6)}
\end{equation}
(and is manifestly independent of the sign of $\epsilon_g$). 

Finally, one can identify the common $O(6)$ subgroup of $\GL(6,\bbR)$
and $O(6)\times O(6)$. From this point of view $\theta_1$ and
$\theta_2$ are spinors of the same $\Spin(6)$ group and
$\chis$ is a bispinor. It is natural to represent $\chis$ as
\begin{equation}
\label{bispin}
   \chis = \tfrac{1}{4}\theta_1\theta_2^t(1-\gamma_{(6)}) 
     = \sum_p\frac{1}{8p!}
       \chi_{m_1\dots m_p}\gamma^{m_1\dots m_p} , 
\end{equation}
where the component forms are given by 
\begin{equation}
   \chi_{m_1\dots m_p} = \tr(\chi\gamma_{m_p\dots m_1}) \in
\Lambda^pF\ .
\end{equation}
The additional factor of $1-\gamma_{(6)}$ is included so that the
induced Clifford action on the forms $\chi_p$ is that given
in~\eqref{Caction}. In terms of this representation~\eqref{bispin} the
spinor bilinear is given by  
\begin{equation}
   \psis^t\cdot\chis = - 8 \tr (\psis^t\chis) 
\end{equation}
where in this representation one has 
\begin{equation}
   \psis^t = \gamma_{(6)}C\psis^TC^{-1} , 
\end{equation}
which follows directly from~\eqref{interdecomp}
and~\eqref{bispin}. Similarly given the expression~\eqref{interdecomp}
for the intertwiner $\tilde{\mathcal{D}}$, we have 
\begin{equation}
\label{Drep}
   \chis^{\tilde{c}} 
     = \tilde{\mathcal{D}}\chi^* 
     = \gamma_{(6)}D\chis^*D^{-1}\gamma_{(6)}^{-1}\ .
\end{equation}
In terms of the component forms $\chi^{\tilde{c}}_p=\chi_p^*$.  

Let us finish by considering chiral spinors $\chis^\pm\in S^\pm$ in
the representation~\eqref{bispin}. First we note that in this case the
Clifford action can be written as 
\begin{equation}
   (V^\Sigma\Gamma_\Sigma)\chis^\pm 
      = \tfrac{1}{2}[ x^m\gamma_m , \chis^\pm ]_\mp 
         + \tfrac{1}{2}[ \xi_m\gamma^m , \chis^\pm ]_\pm . 
\end{equation}
Next, given the chirality operator~\eqref{chdef}, we see that real
chiral spinors can be written as 
\begin{equation}
\label{real-ch}
   \chi_\epsilon^\pm = \zeta_+\bar{\zeta}'_\pm 
      \pm  \zeta_-\bar{\zeta}'_\mp \ ,
\end{equation}
where $\zeta_\pm$ and $\zeta'_\pm$ are chiral $\Spin(6)$
spinors. Note that as such they are eigenspinors of $1-\gamma_{(6)}$
and comparing with~\eqref{bispin} we see this form is compatible with
$\zeta_\pm$ and $\zeta'_\pm$ being sections of the two spin bundles
$S_1$ and $S_2$ respectively. Note that the sign between the two terms
in~\eqref{real-ch} comes from the reality condition
defined using~\eqref{Drep}. 

In the main text we are interested in a pair of complex chiral
$O(6,6)$ spinors given in the representation~\eqref{bispin} by 
\begin{equation}
   \Phi_0^+ = \eta^1_+\bar\eta^2_+ , \qquad
   \Phi_0^- = \eta^1_+\bar\eta^2_- .
\end{equation}
Note, that, in this case we have 
\begin{equation}
\label{conjugates}
   (\Phi_0^+)^{\tilde{c}} = \tilde{\mathcal{D}}(\Phi_0^+)^*
      = \eta^1_-\bar\eta^2_- , \qquad
   (\Phi_0^-)^{\tilde{c}} = \tilde{\mathcal{D}}(\Phi_0^-)^* 
      = - \eta^1_-\bar\eta^2_+ .
\end{equation}
By a slight abuse of notation, in the main text we denote
$(\Phi_0^\pm)^{\tilde{c}}$ by $\bar{\Phi}_0^\pm$.  Note that we also
have
\begin{equation}
\label{transposes}
   (\Phi_0^+)^t = - \ii\eta^2_-\bar\eta^1_- , \qquad
   (\Phi_0^-)^t = \ii\eta^2_+\bar\eta^1_- .
\end{equation}
%


\section{Generic truncation}
\label{app:expand}


In this appendix we discuss the general conditions on mode truncations
of the infinite tower of Kaluza--Klein states on $M_6$. In particular,
we give the conditions such that there is a local special K\"ahler
metric on the moduli space truncated modes, which is inherited from
the local special K\"ahler geometry of the untruncated theory. A
special case of such a truncation, is the expansion in terms of
harmonic modes on a Calabi--Yau manifold.\footnote{A discussion of the
  truncation conditions in the particular case of an $\SU(3)$
  structure also appeared very recently in~\cite{mkp} and appears to
  be in agreement with the analysis given here.}  

The section is divided as follows. We first recall the definition of
(local) special K\"ahler geometry following the approach
of~\cite{freed}. We then review how this geometry is realised in the
untruncated theory and finally derive the conditions for a special
K\"ahler geometry on the truncated theory. 

\subsection{Special K\"ahler geometry}

There are many different ways to define a rigid or local special
K\"ahler geometry. One is as follows~\cite{freed}. Let $U$ be a
$2d$-dimensional K\"ahler manifold with K\"ahler form $\omega$ and
complex structure $J$. A rigid special K\"ahler structure on $U$ is a
flat torsion-free connection $\nabla$ satisfying
\begin{equation}
   \nabla_i\omega_{jk} = 0 , \qquad
   \nabla_{[i}J^k{}_{j]}=0 .
\end{equation}
The first condition is equivalent to the
statement that one can find coordinates $u^i$ whose transition
functions are of the form
\begin{equation}
\label{flatcoord}
   u^i = S^i{}_j u'^j + a^i , 
\end{equation}
where $S\in\Symp(2d,\bbR)$ is a constant symplectic transformation
and $a\in\bbR^{2d}$. In these coordinates $\nabla_i=\partial_i$. The
second condition means that locally one can introduce a vector
$\hat{u}=\hat{u}^i\partial_i$ such that, in these coordinates, 
\begin{equation}
\label{hatu}
   J^i{}_j = - \partial_j\hat{u}^i . 
\end{equation}
Furthermore since the metric $g_{ij}=\omega_{ik}J^k{}_j$ is symmetric
we have locally 
\begin{equation}
\label{K-def}
   \hat{u}^i = - (\omega^{-1})^{ij} \partial_j K 
\end{equation}
for some real function $K$. In addition, it is easy to see that $K$ is
actually the K\"ahler potential. 

One can introduce special complex coordinates as follows. Given the
coordinates $u^i$, locally one can define a vector field
$u=u^i\partial_i$ and hence a local holomorphic vector field  
\begin{equation}
\label{z-def}
   \zeta = \tfrac{1}{2} \left(u + \ii \hat{u} \right).
\end{equation}
From~\eqref{flatcoord} and~\eqref{hatu} we see that $\zeta$ is
unique up to a shift by a constant complex vector. Furthermore
\begin{equation}
   K_{\text{rigid}} = \ii \omega(\zeta,\bar{\zeta}) .
\end{equation}
By making a symplectic transformation one can always choose Darboux
coordinates $u^i=(x^I,y_I)$ with $I=1,\dots,d$ such that 
\begin{equation}
   \omega = \dd x^I \wedge \dd y_I .
\end{equation}
In this basis one can write $\zeta$ as 
\begin{equation}
   \zeta = Z^I \frac{\del}{\del x^I} 
      - \mathcal{F}_I \frac{\del}{\del y_I} . 
\end{equation}
The functions $Z^I$ are special complex coordinates on the special
K\"ahler manifold and the holomorphic functions $\mathcal{F}_I$ are locally
given in terms of a prepotential $\mathcal{F}(Z)$, by
$\mathcal{F}_I=\del\mathcal{F}/\del Z^I$. 

A local special K\"ahler manifold can be viewed as a quotient of a
rigid special K\"ahler manifold. Suppose $U$ is a $2d+2$ dimensional
rigid special K\"ahler manifold such that one can find a globally
defined holomorphic vector field $\zeta$ of the form~\eqref{z-def}
such that $\im\zeta$ is a Killing vector field and the orbits of $\zeta$
define $U$ as a $\bbC^*$ fibration over a base $V$. The space $V$ is
then a special K\"ahler manifold and the metric induced on $V$ by
taking the quotient by the $\bbC^*$ action is a local special K\"ahler
metric. The special coordinates $Z^I$ become projective special
coordinates on $V$. The $\bbC^*$ symmetry implies that the prepotential
$\mathcal{F}(Z)$ is homogeneous of degree two. The K\"ahler potential
on $V$ is given by 
\begin{equation}
   K = - \ln \ii \omega(\zeta,\bar{\zeta}) .
\end{equation}

The moduli space of Calabi-Yau manifolds is a product of two special
geometries spanned by the deformations of the K\"ahler form and the
deformations of the complex structure \cite{Strominger}.

\subsection{Truncation conditions}
\label{sec:trunc}

\subsubsection*{The untruncated theory}

Let us now review how special K\"ahler manifolds appear in the context
of generalised geometry following~\cite{3form,HitchinHF,GCY} (see
also~\cite{GLW} for a review).  Let $S^\pm(E)$ be the positive and
negative chirality generalised spinor bundles discussed in
section~\ref{structure} and $S^\pm_p(E)$ be the fibre at a point $p\in
M^{9,1}$. One then considers an open subset $\mathcal{S}^\pm_p\subset
S_p^\pm(E)$ of so-called stable spinors. These are the spinors with
stabilizer group $\SU(3,3)$. One finds that $U$ is an open orbit under
$O(6,6)$.   

Hitchin then shows that there is a natural local special K\"ahler
metric on $\mathcal{S}^\pm_p$. The construction is as follows. Since
$S^\pm_p(E)$ is a vector space one can identify $T\mathcal{S}^\pm_p$
with $S^\pm_p(E)$ and define the symplectic structure $\omega$ in
terms of the spinor bilinear~\eqref{mukai}, that is, for
$\psi,\chis\in S^\pm_p(E)\simeq T\mathcal{S}^\pm_p$, 
\begin{equation}
   \omega(\psis,\chis) = \psis^t \cdot \chis .
\end{equation}
One then chooses natural coordinates $\chis^i$ which are just the
components of the spinor $\chis\in \mathcal{S}^\pm_p$.  Then by definition
$\nabla_i\omega_{jk}=0$ with $\nabla_i=\der/\der\chis^i$. 

The complex structure is defined by the real function $K_\text{rigid}$
via~\eqref{hatu} and~\eqref{K-def}. On $\mathcal{S}^\pm_p$ it is given
by Hitchin function  
\begin{equation}
   K_\text{rigid} = H_\epsilon(\chis) . 
\end{equation}
This is a particular $\Spin(6,6)$ invariant homogeneous function of
degree two. In the notation of~\cite{GLW} the holomorphic vector field
$\zeta$ is given by 
\begin{equation}
   \Phi_\epsilon^\pm 
      = \tfrac{1}{2}\left( \chis + \ii \hat{\chi}_\epsilon \right)
\end{equation}
where $\hat{\chi}^i_\epsilon=-(\omega^{-1})^{ij}(\der
H_\epsilon/\der\chis^j)$, and is precisely the pure spinor
$\Phi_\epsilon^+$ or $\Phi_\epsilon^-$ discussed in
section~\ref{structure} which was used to define an $\SU(3,3)$
structure.  

Finally, the homogeneity of $H_\epsilon$ implies that
$\hat{\chi}_\epsilon$ is a Killing vector field. Furthermore
$\mathcal{S}^\pm_p$ is a $\bbC^*$ fibration, where $\Phi_\epsilon^\pm$
generates the $\bbC^*$ action on the fibres. This implies that the
quotient $\mathcal{S}^\pm_p/\bbC^*$ is a local special K\"ahler
manifold with K\"ahler potential
\begin{equation}
\label{spinH}
   K = - \ln H_\epsilon .
\end{equation}
Note that this implies that the corresponding metric is actually
independent of the choice of volume form which defines the isomorphism
between $S^\pm$ and $\Leo$. These means that the how analysis could
actually be repeated for stable forms $\chi\in\Leo$. In this case, the
symplectic structure gets replaced by the Mukai pairing~\eqref{mukai}
and the Hitchin function becomes a six-form
\begin{equation}
\label{formH}
   \ee^{-K} = H = \ii \mukai{\Phi^\pm}{\bar{\Phi}^\pm}
\end{equation}

Crucially, the local special K\"ahler metric on $V_p$ defined
by~\eqref{spinH} or~\eqref{formH}, is related to the supergravity
action. Specifically in the case of $\SU(3)$ structure it was shown
that the metrics on $\mathcal{S}^\pm_p/\bbC^*$ corresponding to the
two pure spinors $\Phi^\pm$ are related to the corresponding kinetic
terms in the rewriting of type II supergravity.  

\subsubsection*{Defining the truncation}

Now suppose that $M^{9,1}=M^{3,1}\times M^6$ so that $F=TM^6$. In
analogy to keeping only the moduli of a Calabi--Yau manifold we would
like to make a truncation, keeping some finite dimensional subspace of
$\SU(3,3)$ structures $\Phi$ on $E$. More formally let us start by
defining a sub-bundle $\mathcal{S}^\pm\in S^\pm(E)$ of stable spinors (or
the equivalent space of stable odd or even forms). The truncation is
then an embedding map from some finite dimensional space $U$ into the
infinite dimensional space of sections $\scs{\mathcal{S}^\pm}$ 
\begin{equation}
\label{trunc}
   \tn : U \to \scs{\mathcal{S}^\pm} . 
\end{equation}
In the case of a Calabi--Yau manifold, $U$ is the odd or even
cohomology and $\tn$ identifies harmonic forms with elements in $U$. 
For the truncation to be supersymmetric, we require that the special
K\"ahler geometry on the fibres $\mathcal{S}^\pm_p$ induces a special
K\"ahler metric on $U$. The purpose of this section is to find the
constraints on the map $\tn$ such that this is true. 

The first requirement is that $U$ is a complex manifold. We have
already seen that there is a natural complex structure on each fibre
$\mathcal{S}^\pm_p$. Hence there is a complex structure $\mathcal{J}$ on
$\scs{\mathcal{S}^\pm}$. This will descend to a complex structure on $U$
if the embedding $\tn$ is holomorphic. Specifically, recall that $\tn$
induces the usual push-forward map $\tn_*:TU\to T\mathcal{S}^\pm$ on
vectors. We then define the complex structure $J$ on $U$ by requiring
it to be compatible with the complex structure $\mathcal{J}$ on
$\scs{\mathcal{S}^\pm}$, that is $\tn_*J=\mathcal{J}\tn_*$. Explicitly
suppose $u^i$ are coordinates on $U$. In general we can write the
push-forward of a vector $t\in TU$ as 
\begin{equation}
   t = t^i\der_i \mapsto \tn_*t = t^i \Sigma_i(u) 
\end{equation}
where $\Sigma_i(u)=\der_i\tn$ form a basis\footnote{In the main text, we use
the notation $\Ae_{\hA}$ for the basis of even forms in $T\mathcal{S}^+$, 
and $\Ao_{\hI}$ for the basis of odd forms in $T\mathcal{S}^-$.} for
the image of $TU$ in $T\mathcal{S}^\pm$. In the special case of a
Calabi--Yau manifold, $\Sigma_i(u)$ are harmonic forms. The complex
structure $J$ is then related to $\mathcal{J}$ by
\begin{equation}
\label{J-def}
   \mathcal{J}\Sigma_i = J^j{}_i \Sigma_j .
\end{equation}
In other words the image of $\Sigma_i$ under $\mathcal{J}$ can still
be expanded in the basis $\Sigma_i$. In the context of a Calabi--Yau
manifold that action of $\mathcal{J}$ corresponds to taking the Hodge
dual. The condition~\eqref{J-def} then states that the Hodge dual of a
harmonic form is itself harmonic. 

We now turn to the symplectic structure on $U$. We have seen that the
spinor bilinear (or equivalently the Mukai pairing) defines a
symplectic structure on each fibre $\mathcal{S}_p$. We can define a
bilinear on $\scs{T\mathcal{S}}$ simply by integrating over
$M^6$. Using $\tn_*$ we can then define a bilinear $\omega$ on $TU$ by
\begin{equation}
\label{o-def}
   \omega(s,t) = \int_{M^6} \mukai{\tn_*s}{\tn_*t} .
\end{equation}
In components we have 
\begin{equation}
   \omega_{ij} = \int_{M^6} \mukai{\Sigma_i}{\Sigma_j} .
\end{equation}
To be a symplectic structure we require that $\omega$ is
non-degenerate. Using the K\"ahler structure on $\mathcal{S}_p$, it is
then by construction compatible with $J$. 

The next requirement is that $(\omega,J)$ is special K\"ahler. This
means first that we can choose coordinate $u^i$ such that
$\der_i\omega_{jk}=0$ or equivalently 
\begin{equation}
\label{sK-cond}
   \int_{M^6} \mukai{\Sigma_j}{\der_i\Sigma_k} = 0 .
\end{equation}
Again, the special K\"ahler structure on $\mathcal{S}_p$ then implies
that $\der_{[i}J^k{}_{j]}=0$ and hence there is a rigid special
K\"ahler metric on $U$. 

Finally, of course, we actually want a local special K\"ahler
metric, and hence some natural $\bbC^*$ action on $U$. Again, we have
such an action on $\mathcal{S}_p$ generated by the holomorphic vector
$\Phi$ and hence a $\bbC^*$ action on $\scs{\mathcal{S}}$. Thus the
natural requirement is that this induces a $\bbC^*$ action on $U$. In
other words the holomorphic vector $\zeta\in TU$ of the
form~\eqref{z-def} which defines the rigid special K\"ahler structure
on $U$ satisfies $\tn_*\zeta=\Phi$. This means that, on a coordinate
patch $u^i$ the map $\tn$ is realised by 
\begin{equation}
   u^i \mapsto u^i\Sigma_i .
\end{equation}
Since we also have $\Sigma_i=\der_i\tn$ this requires that
$u^i\der_j\Sigma_i=0$ or equivalently
\begin{equation}
\label{local-cond}
   u^i\der_i\Sigma_j = 0 ,
\end{equation}
that is, the basis forms $\Sigma_i$ are homogeneous of degree zero. If
this is satisfied, then there is a local special K\"ahler metric on
$V=U/\bbC^*$. Furthermore, it is easy to show that the K\"ahler
potential on $V$ is given by 
\begin{equation}
   K = - \ln \int_{M^6} H 
      = - \ln \ii \int_{M_6}\mukai{\Phi^\pm}{\bar{\Phi}^\pm}
\end{equation}
where $H$ is the Hitchin function defined using the Mukai pairing. 

Finally, it is convenient to rewrite these expressions in terms of
Darboux coordinates $u^i=(x^I,y_I)$ with $I=0,1,\dots,d$ such that
$\omega=\dd x^I\wedge\dd y_I$. Distinguishing between the odd and even
cases we have the corresponding bases
\begin{equation}
   \Ae= \{\omega_A,\tilde{\omega}^B\} , \qquad \qquad 
   \Ao = \{\alpha_I,\beta^J\} 
\end{equation}
such that 
\begin{equation}
   \int_{M^6} \mukai{\alpha_I}{\beta^J} = \delta_I{}^J ,
\end{equation}
and $\int_{M_6}\mukai{\alpha_I}{\alpha_J}=
\int_{M_6}\mukai{\beta^I}{\beta^J}=0$, together with 
\begin{equation}
   \int_{M^6} \mukai{\omega_A}{\tilde{\omega}^B} = \delta_A{}^B ,
\end{equation}
and $\int_{M^6}\mukai{\omega_A}{\omega_B}=
\int_{M^6}\mukai{\tilde{\omega}^A}{\tilde{\omega}^B}=0$. 

We can then introduce holomorphic coordinates $Z^I$ (or $X^A$) and a
prepotential $\mathcal{F}$ (or $F$) such that 
\begin{equation}
\begin{aligned}
   \Phi^+ &= X^I \omega_A - F_A \tilde{\omega}^A , \\ 
   \Phi^- &= Z^I \alpha_I - \mathcal{F}_I \beta^I .
\end{aligned}
\end{equation}
%



\end{document}